\documentclass[a4paper,fleqn]{cas-sc}

\usepackage{lipsum}
\usepackage[utf8]{inputenc}
\usepackage[numbers]{natbib}

\xspaceaddexceptions{]}

\def\be{\begin{equation}}
	\def\ee{\end{equation}}
\newcommand{\bel}[1]{\begin{eqnarray}\label{#1}}
	\newcommand{\eel}{\end{eqnarray}}
\def\barr{\begin{array}}
	\def\earr{\end{array}}
\def\beq{\begin{eqnarray}}
	\def\eeq{\end{eqnarray}}
\def\bfig{\begin{figure}}
	\def\efig{\end{figure}}
\def\lt{\left}
\def\rt{\right}
\newcommand{\nn}{\nonumber}
\newcommand{\f}[2]{\frac{#1}{#2}}
\newcommand{\onehalf}{{\nicefrac{1}{2}}}

\newcommand{\p}{\partial}

\newcommand{\tr}{{\rm tr}}
\newcommand{\rf}[1]{Eq.~(\ref{#1})}

\newcommand{\rfn}[1]{(\ref{#1})}

\def\a{\alpha}
\def\b{\beta}
\def\g{\gamma}
\def\d{\delta}

\def\c{\chi}
 
\def\LR{\left(} 
\def\RR{\right)}
\def\LS{\left[} 
\def\RS{\right]}

\def\HP{\hphantom{\alpha}} 

\newcommand{\sh}[1]{\sinh#1}
\newcommand{\ch}[1]{\cosh#1}

\def\half{\frac{1}{2}}

\def\GLW{{\rm GLW}}

\def\nU{n_{(0)}}
\def\eU{\varepsilon_{(0)}}
\def\PU{P_{(0)}}
\def\sU{s_{(0)}}

\def\wP{w_{}}

\def\Cv{{\boldsymbol C}}

\usepackage{amssymb}

\usepackage{graphicx}
\usepackage{amsmath}
\usepackage{subfigure}
\usepackage{bbold}
\usepackage{yfonts}
\usepackage{placeins}
\usepackage{bm} 
\usepackage{nicefrac}
\usepackage{slashed}

\newcommand{\bea}{\begin{eqnarray}}
\newcommand{\eea}{\end{eqnarray}}

\def\LB{\left(}
\def\RB{\right)}
\def\LSB{\left[}
\def\RSB{\right]}

\newcommand{\EQ}[1]{Eq.~(\ref{#1})}
\newcommand{\EQn}[1]{(\ref{#1})}

\newcommand{\EQS}[1]{Eqs.~(\ref{#1})}

\newcommand{\EQSTWO}[2]{Eqs.~(\ref{#1})~and~(\ref{#2})}
\newcommand{\EQSTWOn}[2]{(\ref{#1})~and~(\ref{#2})}
\newcommand{\EQSM}[2]{Eqs.~(\ref{#1})--(\ref{#2})}
\newcommand{\EQSMn}[2]{(\ref{#1})--(\ref{#2})}

\newcommand{\SEC}[1]{Sec.~\ref{#1}}

\newcommand{\CIT}[1]{Ref.~\cite{#1}} 
\newcommand{\CITS}[1]{Refs.~\cite{#1}} 
\newcommand{\CITn}[1]{\cite{#1}}

\def\spin{\,\textgoth{s:}}
\def\spinl{|{\boldsymbol s}_*|}

\def\Lg{\,L}

\newcommand{\bv}{{\boldsymbol b}} 
\newcommand{\ev}{{\boldsymbol e}}

\newcommand{\vv}{{\boldsymbol v}}
\newcommand{\pv}{{\boldsymbol p}}

\newcommand{\Pv}{{\boldsymbol P}}

\newcommand{\piv}{{\boldsymbol \pi}}
\newcommand{\sv}{{\boldsymbol s}}
\def\sigv{{\boldsymbol \sigma}}

\def\pmU{p^\mu}

\newcommand{\ii}{i}
\newcommand{\di}{{\rm d}} 

\newcommand{\trt}{{\rm tr_2}}
\newcommand{\trf}{{\rm tr_4}}

\def\wpsi{{\widehat{\psi}}}
\def\wPhi{{\widehat{\Phi}}}

\def\bmL{\beta_\mu}

\def\umU{u^\mu}

\def\omnL{\omega_{\mu\nu}}
\def\omnU{\omega^{\mu\nu}}
\def\onmL{\omega_{\nu\mu}}

\def\oabL{\omega_{\alpha\beta}}

\def\omnLD{{\tilde \omega}_{\mu\nu}}
\def\omnUD{\tilde {\omega}^{\mu\nu}}

\def\epsUabgd{\epsilon^{\alpha \beta \gamma \delta}}

\def\epsLmnab{\epsilon_{\mu\nu\alpha\beta}}

\def\epsUmnrs{\epsilon^{\mu\nu\rho \sigma}}

\def\gfive{\gamma_5}

\def\wP{{\widehat{P}}}
\def\TmnU{T^{\mu\nu}}
\def\TnmU{T^{\nu\mu}}
\def\TpmnU{T^{\prime\, \mu\nu}}

\def\TmnUB{T^{\mu\nu}_{\rm Bel}}
\def\TnmUB{T^{\nu\mu}_{\rm Bel}}

\def\oTmnU{{\widehat T}^{\mu\nu}}
\def\oTmlU{{\widehat T}^{\mu\lambda}}
\def\oTpmnU{{\widehat T}^{\,\prime\, \mu\nu}}
\def\oTmnUB{{\widehat T}^{\mu\nu}_{\rm Bel}}
\def\oTmnUc{{\widehat T}^{\mu\nu}_{\rm can}}

\def\oTmlUc{{\widehat T}^{\mu\lambda}_{\rm can}}
\def\oTnlUc{{\widehat T}^{\nu\lambda}_{\rm can}}
\def\oTlnUc{{\widehat T}^{\lambda\nu}_{\rm can}}

\def\SmnU{{\Sigma}^{\mu\nu}}
\def\SmnL{{\Sigma}_{\mu\nu}}
\def\SabU{{\Sigma}^{\alpha\beta}} 
\def\S0iU{{\Sigma}^{0i}}

\def\wJ{{\widehat{J}}}

\def\JmlnU{J^{\mu, \lambda\nu}}

\def\JlmnUB{J^{\lambda, \mu\nu}_{\rm Bel}}

\def\oJmlnU{{\widehat J}^{\mu, \lambda\nu}}
\def\oJpmlnU{{\widehat J}^{\,\prime\,\mu, \lambda\nu}}

\def\oJmlnUc{{\widehat J}^{\mu, \lambda\nu}_{\rm can}}

\def\oLc{{\widehat L}_{\rm can}}

\def\oLmlnUc{{\widehat L}^{\mu, \lambda\nu}_{\rm can}}

\def\oSc{{\widehat S}_{\rm can}}
\def\SlmnU{S^{\lambda, \mu\nu}}
\def\SmlnU{S^{\mu, \lambda\nu}}
\def\SlmnUB{{S^{\lambda, \mu\nu}_{\rm Bel}}}

\def\oSlmnU{{{\widehat S}^{\lambda, \mu\nu}}}
\def\oSmlnU{{{\widehat S}^{\mu, \lambda\nu}}}
\def\oSplmnU{{{\widehat S}^{\,\prime\, \lambda, \mu\nu}}}
\def\oSlmnUc{{{\widehat S}^{\lambda, \mu\nu}_{\rm can}}}
\def\oSmlnUc{{{\widehat S}^{\mu, \lambda\nu}_{\rm can}}}
\def\oSnlmUc{{{\widehat S}^{\nu, \lambda\mu}_{\rm can}}}

\def\wj{{\widehat j}}

\def\n0{n_{(0)}}
\def\e0{\varepsilon_{(0)}}
\def\P0{P_{(0)}}

\def\TmnU{T^{\mu\nu}}                      

\def\rhoLEQ{{\widehat{\rho}}_{\rm \small LEQ}}
\def\rhoGEQ{{\widehat{\rho}}_{\rm \small GEQ}}

\def\fplusrsxp{f^+_{rs}(x,p)}

\def\fminusrsxp{f^-_{rs}(x,p)}

\def\ubarrp{{\bar u}_r(p)}

\def\usp{u_s(p)}
\def\urp{u_r(p)}

\def\vbarrp{{\bar v}_r(p)}
\def\vbarsp{{\bar v}_s(p)}
\def\vsp{v_s(p)}
\def\vrp{v_r(p)}

\def\Wpmxk{{\cal W}^{\pm}(x,k)}
\def\Wxk{{\cal W}(x,k)}
\def\Weqxk{{\cal W}_{\rm eq}(x,k)}
\def\Weqpxk{{\cal W}^{+}_{\rm eq}(x,k)}
\def\Weqmxk{{\cal W}^{-}_{\rm eq}(x,k)}
\def\Weqpmxk{{\cal W}^{\pm}_{\rm eq}(x,k)}

\def\Feqpmxk{{\cal F}^{\pm}_{\rm eq}(x,k)}

\def\Peqpmxk{{\cal P}^{\pm}_{\rm eq}(x,k)}

\begin{document}
\let\WriteBookmarks\relax
\def\floatpagepagefraction{1}
\def\textpagefraction{.001}
\shorttitle{Relativistic hydrodynamics for spin-polarized fluids}
\shortauthors{Florkowski et al.}

\title [mode = title]{Relativistic hydrodynamics for spin-polarized fluids}                      
\tnotemark[1]
 
\tnotetext[1]{This work was supported in part by the Polish National Science Center Grants   No. 2016/23/B/ST2/00717 and No. 2018/30/E/ST2/00432.} 

\author[1]{Wojciech Florkowski}[role=Author] 
\ead{Wojciech.Florkowski@uj.edu.pl} 
\address[1]{M. Smoluchowski Institute of Physics, Jagiellonian University,  PL-30-348 Krak\'ow, Poland}

\author[2]{Radoslaw Ryblewski}[role=Author]
\ead{Radoslaw.Ryblewski@ifj.edu.pl} 
\author[2]{Avdhesh Kumar}[role=Author]
\ead{Avdhesh.Kumar@ifj.edu.pl} 
\address[2]{Institute of Nuclear Physics, Polish Academy of Sciences, 31-342 Krak\'ow, Poland }
  
\begin{abstract}[S U M M A R Y]
Recent progress in the formulation of relativistic hydrodynamics for particles with spin one-half is reviewed. We start with general arguments advising introduction of a tensor spin chemical potential that plays a role of the Lagrange multiplier coupled to the spin angular momentum. Then, we turn to a discussion of spin-dependent distribution functions that have been recently proposed to construct a hydrodynamic framework including spin and serve as a tool in phenomenological studies of hadron polarization. Distribution functions of this type are subsequently used to construct the equilibrium Wigner functions that are employed in the semi-classical kinetic equation. The semi-classical expansion elucidates several aspects of the hydrodynamic approach, in particular, shows the ways in which different possible versions of hydrodynamics with spin can be connected by  pseudo-gauge transformations. These results point out at using the de~Groot - van~Leeuwen - van~Weert versions of the energy-momentum and spin tensors as the most natural and complete physical variables. Finally, a totally new method is proposed to design hydrodynamics with spin, which is based on the classical treatment of spin degrees of freedom. Interestingly, for small values of the spin chemical potential the new scheme brings the results that coincide with those obtained before. The classical approach also helps us to resolve problems connected with the normalization of the spin polarization three-vector. In addition, it clarifies the role of the Pauli-Luba\'nski vector and the entropy current conservation. We close our review with several general comments presenting possible future developments of the discussed frameworks. 
\end{abstract}

\maketitle

\tableofcontents

\section{Introduction}

  Non-central heavy-ion collisions   at low and intermediate energies  create strongly-interacting systems with large global angular momenta. This may produce spin polarization of the hot and dense matter in a way similar to the magnetomechanical effects of Einstein and de Haas~\cite{dehaas:1915} and Barnett~\cite{RevModPhys.7.129}. Interestingly, the longitudinal polarization of the ${\bar \Lambda}$ hyperons was discussed as early as in 1980s by Jacob and Rafelski in connection with the formation of a  quark-gluon plasma \cite{Jacob:1987sj}. However,  the first heavy-ion experiments that measured the $\Lambda$ spin polarization in Dubna \cite{Anikina:1984cu} and at CERN  \cite{Bartke:1990cn} reported negative results. 

The first predictions on global polarization of produced particles, based on a perturbative-QCD inspired model and spin-orbit interactions, were proposed in Refs.~\cite{Liang:2004ph,Liang:2004xn} and \cite{Betz:2007kg}, see also \cite{Voloshin:2004ha}. These works predicted a rather substantial polarization effect, of the order of 10\%, which was not confirmed by the STAR result of 2007 \cite{Abelev:2007zk}. Subsequently, the idea of a non-vanishing global polarization was proposed, which referred to local thermodynamic equilibrium of the spin degrees of freedom \cite{Becattini:2007nd,Becattini:2007sr,Becattini:2013fla,Becattini:2013vja,Becattini:2016gvu}, which indicated that the polarization could be much lower than the QCD-inspired prediction of 2005 and of the order of 1\%. These updated calculations drew the attention of the experimentalists and convinced them to attempt a new measurement with a larger statistics in the years 2015--2017. This eventually led to  the first positive measurement of the global  $\Lambda$--hyperon spin polarization by STAR~\cite{STAR:2017ckg,Adam:2018ivw} being in quantitative agreement with the theoretical predictions \cite{Karpenko:2016jyx,Li:2017slc,Xie:2017upb,Sun:2017xhx}. 

 Theoretical studies analyzing the spin polarization and vorticity formation in heavy-ion collisions include several highly debated topics: the importance of the spin-orbit coupling~\cite{Gao:2007bc,Chen:2008wh}, global equilibrium with a rigid rotation~\cite{Becattini:2009wh,Becattini:2012tc,Becattini:2015nva,Hayata:2015lga}, kinetic models of spin dynamics~\cite{Gao:2012ix,Chen:2012ca,Fang:2016vpj,Fang:2016uds}, anomalous hydrodynamics~\cite{Son:2009tf,Kharzeev:2010gr}, the Lagrangian formulation of hydrodynamics~\cite{Montenegro:2017rbu,Montenegro:2017lvf,Montenegro:2018bcf,Montenegro:2019tku} and hydrodynamic treatment of the spin tensor using holographic techniques \cite{Gallegos}.  

Despite these efforts, the present works say little about the changes of spin polarization during the heavy-ion collision process, in particular, if the latter is described with the help of fluid dynamics which has become now the basic ingredient of heavy-ion  models (the latest  developments within relativistic hydrodynamics have recently been reviewed in \CITS{Florkowski:2017olj,Romatschke:2017ejr}). This is quite surprising, since the studies of fluids with spin have a long history initiated in 1930s \cite{Weyssenhoff:1947iua,PhysRev.109.1882,Halbwachs:1960aa}. Recent works have mainly contributed to our understanding of global equilibrium states which exhibit interesting features of vorticity-spin alignment \cite{Becattini:2009wh,Becattini:2015nva}, and polarization effects present at the final, kinetic freeze-out stage of collisions \cite{Becattini:2013fla,Becattini:2013vja,Fang:2016vpj}. This situation has changed very recently with the formulation of a hydrodynamic framework \cite{Florkowski:2017ruc,Florkowski:2017dyn} that allows for studies of space-time evolution of the spin polarization~\cite{Florkowski:2018ual,Florkowski:2018amv}.

The formulation of hydrodynamics with spin proposed in Refs.~\cite{Florkowski:2017ruc,Florkowski:2017dyn} is based, however, on a particular choice of the forms of energy-momentum and spin tensors. More recent works have clarified the use of different forms of such tensors by the analysis of their physical significance and by explicit constructions of the pseudo-gauge transformations that relate different frameworks~\CITn{Becattini:2018duy,Florkowski:2018ahw}. In the first sections of this work we review some of the most important findings of~\CITS{Becattini:2018duy,Florkowski:2018ahw} and introduce the concept of a spin chemical potential. In the following sections we advocate future applications of the de Groot - van Leeuwen - van Weert (GLW) formalism~\CITn{DeGroot:1980dk}. A totally new aspect of our presentation is the introduction of a framework that treats spin classically~\CITn{Weert:1970aa}. We show that this approach agrees with the quantum GLW results for small values of the spin chemical potential. Moreover, this framework: i)  indicates how one can avoid problems with the normalization of the average polarization three-vector for large values of the spin potential, ii) helps to define microscopic conditions that validate the use of the proposed equilibrium functions, and iii) can be used to define entropy current and prove its conservation within the perfect-fluid approach with spin. The results obtained with the classical treatment of spin can be helpful for future construction of a  dynamic quantum description based on the spin density matrix and not restricted to small values of the spin chemical potential.
  
Last but not least, we stress that our presentation uses the tools and concepts of kinetic theory. This allows to study the mean spin polarization of constituent particles (or quasi-particles) and to determine its space-time evolution. Such a picture might be not valid for strongly interacting quark-gluon plasma, especially close to the deconfinement phase transition,  where the very concept of quasi-particles breaks down.

\bigskip
{\it Conventions and notation:} We use the following conventions and notation for the metric tensor, the four-dimensional Levi-Civita tensor, and the scalar product in flat Minkowski space: $g_{\mu\nu} = \hbox{diag}(+1,-1,-1,-1)$, $\epsilon^{0123} = -\epsilon_{0123} = 1$, $a^\mu b_\mu= g_{\mu \nu} a^\mu b^\nu$. Three-vectors are shown in bold font and a dot is used to denote the scalar product of both four- and three-vectors, hence, $a^\mu b_\mu = a \cdot b = a^0 b^0 - \boldsymbol{a} \cdot  \boldsymbol{b}$. The symbol $\boldsymbol{1}$ is used for a 4$\times$4 unit matrix. The traces in spin and spinor spaces are distinguished by using the symbols $\trt$ and $\trf$, respectively. The symbol $\tr$ denotes the trace in the Hilbert space. The Lorentz invariant measure in the momentum space is denoted as $dP$, namely
\bel{eq:dP}
dP = \frac{d^3p}{(2 \pi )^3 E_p},
\eel
where $E_p = \sqrt{m^2 + \pv^2}$ is the on-mass-shell particle energy, and $p^\mu = (E_p, \pv)$. The particle momenta which are not necessarily on the mass shell and appear as arguments of the Wigner functions are denoted by the four-vector~$k^\mu$. The square brackets denote antisymmetrization, $t^{[\mu \nu]} =   \left(t^{\mu\nu} - t^{\nu\mu} \right)/2$. The symbol of tilde is used to denote dual tensors, which are obtained from the rank-two antisymmetric tensors $a_{\mu\nu}$ by contraction with the Levi-Civita symbol and division by a factor of two. For example, 
\bel{eq:dual}
{\tilde a}_{\mu\nu} = \f{1}{2} \epsLmnab \,  a^{\alpha \beta} .
\eel
The inverse transformation is
\bel{eq:dualdual}
a^{\rho \sigma} = -\f{1}{2} \epsilon^{\rho \sigma \mu \nu}
{\tilde a}_{\mu \nu}.
\eel
Throughout the paper we use natural units, $c=\hbar=k_B=1$, except for the parts where we discuss semi-classical expansions and it is important to display the Planck constant $\hbar$ explicitly.

Our considerations are restricted to hydrodynamics of spin-$\onehalf$, massive particles. We distinguish between the particle rest frame PRF (connected with a single particle) and the local fluid rest frame  LFRF (connected with a group of particles forming a fluid element). The quantities defined in PRF are denoted by an asterisk, while unlabeled quantities refer to the laboratory frame LAB. For a particle with four-momentum $p^\mu$ in the laboratory frame, the particle rest frame is obtained by boosting $p^\mu$ by the three-velocity $-\pv/E_p$. The boosts considered in this work are all canonical   (also known as pure)  boosts~\cite{Florkowski:2017dyn,Leader:2001gr}.

\section{Spin chemical potential}

Constructions of the hydrodynamic frameworks rely on the local conservation laws. In standard situations one includes the conservation of energy, linear momentum, baryon number, and possibly of other conserved quantities like electric charge or strangeness~\footnote{In the case where the electric charge conservation is independent from the conservation of the baryon number or if the flavor-changing weak interactions may be neglected.}.  While dealing with particles with spin, it is necessary to include also the conservation of angular momentum, which in this case becomes an independent condition. In the similar way as the conservation of energy, linear momentum and baryon number are connected with the introduction of temperature $T$, fluid velocity $\umU$, and baryon chemical potential $\mu_B$, the conservation of the angular momentum for particles with spin requires introduction of a new spin chemical potential~\CITn{Becattini:2018duy}. As a matter of  fact, the spin chemical potential is not a scalar but an antisymmetric rank-two tensor. 

Inclusion of the conservation of angular momentum in  field-theoretical frameworks is connected with a well-known ambiguity of the localization of energy and spin densities~\CITn{Hehl:1976vr}. For any conserved energy-momentum and spin tensors, $\TmnU$ and $\SlmnU$, it is possible to construct a new pair of such quantities, $\TpmnU$ and $S^{\prime\, \lambda, \mu\nu}$, that are also conserved and yield the same total values of the energy, linear momentum, and angular momentum as the original tensors. The conversion rules that relate such energy-momentum and spin tensors are known as pseudo-gauge transformations. The most known example of such a transformation is the Belinfante construction that starts with the canonical expressions for $\TmnU$ and $\SlmnU$, and ends up with a symmetric energy-momentum tensor, $\TmnUB= \TnmUB$, and a vanishing spin tensor $\SlmnUB=0$. In this case the total angular momentum, $\JlmnUB$, has the form of an orbital one and can be expressed solely by $\TmnUB$.

The formalism of relativistic hydrodynamics with spin necessarily uses the concept of the spin tensor. Hence, the problems connected with the energy and spin localization naturally appear in such an approach. The results presented below shed new light on many of them. In particular, a semi-classical expansion (in powers of $\hbar$)  of the Wigner function shows that the leading-order term of the canonical energy-momentum tensor is in fact symmetric. Hence, it can be used in General Relativity which is a classical theory. Moreover, we construct an explicit form of the pseudo-gauge transformation which connects the canonical framework with the formalism of de Groot, van Leeuwen, and van Weert (GLW). Both the canonical and GLW approaches include spin dynamics, and the results obtained in one framework can be translated to the other one. From this point of view, the Belinfante approach seems to be an effective description that might be valid only in special situations; for example, in the case where the spin dynamics is solely determined by other physical quantities like thermal vorticity. In general, the information about spin cannot be encoded in the form of orbital angular momentum, since the latter can be eliminated by changing the Lorentz frame. We note that similar issues are currently discussed in the context of the nucleon spin~\cite{Leader:2013jra}.

\subsection{Pseudo-gauge transformations}
\label{sect:pseudogauget}
\medskip

As the hydrodynamic frameworks are based on the local conservation laws that are used to define local thermodynamic equilibrium, let us start our discussion with general  considerations about locally conserved quantities. 

In relativistic quantum field theory, according to the Noether theorem, for each continuous symmetry of the action there is a corresponding conserved current. The currents associated with the translational symmetry and the Lorentz symmetry~\footnote{By Lorentz symmetry transformations we understand here Lorentz boosts and rotations.} are the canonical energy-momentum and  angular momentum tensors,
\bea
\oTmnUc &=& \sum_a  \f{\partial \cal L}{\partial(\partial_\mu \wpsi^a)} \partial^\nu \wpsi^a
- g^{\mu\nu} {\cal L},  \label{NoetherT} \\
\oJmlnUc &=& x^\lambda  \oTmnUc - x^\nu  \oTmlUc +\oSmlnUc 
\equiv \oLmlnUc + \oSmlnUc.   \label{NoetherJ}
\eea
Here ${\cal L}$ is the Lagrangian density, the hat symbol denotes operators, $\oLc$ is the orbital part of the total angular momentum, while $\oSc$ is its spin part called the spin tensor and obtained from the expression
\bea
\oSmlnUc =  - \ii \sum_{a,b} \f{\partial \cal L}{\partial(\partial_\mu \wpsi^a)}  D \LB J^{\lambda \nu} \RB^a_b \wpsi^b, \label{SC}
\eea  
where $D^a_b$ is the irreducible representation matrix of the Lorentz group specific for the considered  field. The above tensors fulfill the following equations,
\bea
\p_\mu \oTmnUc &=& 0, \label{Tconserv} \\ 
\p_\mu \oJmlnUc  &=& \oTlnUc - \oTnlUc + \p_\mu \oSmlnUc= 0, \label{Jconserv}
\eea
which also gives
\bea
\p_\mu \oSmlnUc &=& \oTnlUc - \oTlnUc . \label{Sdiv}
\eea
At this point it is important to stress that the canonical energy-momentum tensor is not symmetric, hence the spin part of the angular momentum is not conserved separately. Only the total angular momentum is conserved, which is expressed above by~\EQ{Jconserv}.

It turns out, however, that the energy-momentum and angular momentum tensors are not defined
uniquely. Different pairs can be obtained by either changing the Lagrangian 
density or by means of the so-called pseudo-gauge transformations \CITn{Hehl:1976vr},
\bea
\oTpmnU &=& \oTmnU +\f{1}{2} \partial_\lambda
\left( \wPhi^{\lambda, \mu \nu } - \wPhi^{\mu, \lambda \nu} -  \wPhi^{\nu, \lambda \mu}  \right), \label{psg1} \\
\oSplmnU &=& \oSlmnU -\wPhi^{\lambda,\mu\nu}, \label{psg2} 
\eea
where $\wPhi$ is a rank-three tensor field antisymmetric in the last two indices~\footnote{The pseudo-gauge transformation defined by \EQSTWO{psg1}{psg2} can be still generalized by adding the term $\p_\alpha Z^{\alpha \lambda \mu \nu}$ (with the properties $Z^{\alpha \lambda \mu \nu} = -Z^{\lambda \alpha \mu \nu}$ and $Z^{\alpha \lambda \mu \nu} = -Z^{ \alpha \lambda \nu \mu}$) to the right-hand side of \EQ{psg2} \CITn{Hehl:1976vr}. However, in our considerations it is enough to consider the case $Z^{\alpha \lambda \mu \nu}=0$.}. It is called and below referred to as a superpotential. The newly defined tensors preserve the total energy,  linear momentum, and angular momentum,
\bea
\wP^\nu &=&  \int_\Sigma \di \Sigma_\mu \oTmnU = \int_\Sigma \di \Sigma_\mu \oTpmnU , 
\label{totalP} \\
\wJ^{\lambda\nu} &=& \int_\Sigma \di \Sigma_\mu \oJmlnU = \int_\Sigma \di \Sigma_\mu \oJpmlnU,
\label{totalJ}
\eea
as well as the conservation equations \EQSTWOn{Tconserv}{Jconserv}. In \EQSTWO{totalP}{totalJ} the symbol $d\Sigma_\mu$ specifies the element of a space-like hypersurface. The latter can be determined, for example, by the condition $t=\hbox{const.}$, which implies in this case that $d\Sigma_\mu = (dV,0,0,0)$ with $dV$ being a standard three-volume element.

A special case of the pseudo-gauge transformation is that starting with the canonical  definitions  $\oTmnUc$ and $\oSlmnUc$,  and using the spin tensor itself as a superpotential, i.e., $\wPhi = \oSlmnUc$ \cite{Belinfante:1940}. In this case, the new spin tensor vanishes, while the new energy-momentum tensor has the form
\bea
\oTmnUB = \oTmnUc  +\f{1}{2}  \p_\lambda \LB \oSlmnUc  - \oSmlnUc  -  \oSnlmUc  \RB.  \label{belinf}
\eea
In the following we shall refer to $ \oTmnUB$ defined by \EQ{belinf} as to the Belinfante energy-momentum tensor. 

\subsection{Local-equilibrium density operator}
\label{sect:leqop}
\medskip

The local-equilibrium density operator $\rhoLEQ$ is obtained by maximizing the entropy with the constraints specifying mean densities of the conserved currents over the hyper-surface $\Sigma$~\CITn{Zubarev:1966aa,Zubarev:1979aa,Becattini:2014yxa}. In particular, the projections of the mean energy-momentum tensor and charge current onto the normalized vector perpendicular to $\Sigma$ must be equal to the prescribed values $T^{\mu\nu}_0$ and~$j^{\mu}_0$,
\bea
n_\mu \tr \left(\rhoLEQ \, \oTmnU \right)= n_\mu \TmnU_0, 
\qquad n_\mu \tr \left(\rhoLEQ \, \wj^{\mu}\right) = n_\mu j^{\mu}_0.
\label{entmax1}
\eea
In addition to the energy, linear momentum, and charge densities, one should a priori include the angular 
momentum density among the constraints listed above, namely,
\bea
n_\mu \tr \left( \rhoLEQ \, \oJmlnU \right)  
= n_\mu \tr  \left[ \rhoLEQ \left( x^\lambda \oTmnU -  x^\nu \oTmlU + \oSmlnU \right) \right] 
= n_\mu \JmlnU_0. \nn \\
\label{entmax2}
\eea

If one uses the Belinfante versions of the energy-momentum and spin tensors, the latter vanishes and \EQ{entmax2} becomes redundant --- it is already included in \EQ{entmax1}.  In this case the local-equilibrium density operator reads
\bea
\rhoLEQ = \f{1}{Z} \exp \left[-\int_\Sigma \di \Sigma_\mu \left(  \oTmnUB \beta_\nu 
- \xi \, \wj^\mu \right) \right],
\label{leqBel}
\eea
where $\beta_\nu$ and $\xi$ are the Lagrange multiplier functions, whose meaning is the ratio between the local four-velocity $u^\mu$ and temperature $T$ (a four-temperature vector) and the ratio between local chemical potential $\mu$ and $T$, respectively.

On the contrary, if one works with the canonical tensors, the spin tensor does not vanish and \EQ{entmax2} contains a non-trivial part
\bea
n_\mu \tr \left(\rhoLEQ \,  \oSmlnU \right) = n_\mu \SmlnU_0.  \label{entmax3}
\eea
Since this is an independent constraint, one has to introduce an antisymmetric tensor field $\omega_{\lambda\nu}$. It is dubbed the {\it spin chemical potential} or the {\it spin polarization tensor}~\footnote{Below we use more often the name {\it spin polarization tensor}, as it has been used in our earlier publications and we try to avoid proliferation of names. Moreover, one may try to keep the name ``spin chemical potential'' for the tensor $\Omega_{\mu\nu} = T\,\omega_{\mu\nu} $, in analogy with the notation used for the charge chemical potential which is defined as $\mu = T\,\xi$. }. In analogy with the variable $\xi$, the components of $\omega$ play a role of Lagrange multipliers coupled to the spin tensor. Including \EQ{entmax3} in the construction of local equilibrium leads to the form
\bea
\rhoLEQ = \f{1}{Z} \exp \left[-\int_\Sigma \di \Sigma_\mu 
\LB \oTmnUc \beta_\nu  - \f{1}{2} \omega_{\lambda\nu} \oSmlnUc - \xi \wj^\mu   \RB \right].
\label{leqcan}
\eea

It is interesting now to compare \EQSTWO{leqBel}{leqcan}. Substituting \EQ{belinf} into \EQ{leqcan} and using the property that the canonical spin tensor is invariant under cyclic changes of the indices, we find
\bea
\rhoLEQ = \f{1}{Z} \exp \left[-\int_\Sigma \di \Sigma_\mu \left( \oTmnUB \beta_\nu
- \f{1}{2} (\omega_{\lambda\nu} -\varpi_{\lambda\nu})   \oSmlnU  - \xi \wj^\mu  \right)  \right], 
\label{leqcan1}
\eea
where $\varpi_{\lambda\nu}$ is the thermal vorticity~\footnote{Different possible definitions of the relativistic vorticity and their physical significance is discussed, for example,  in~\CIT{Becattini:2015ska}.}
\bea
\varpi_{\lambda\nu} = -\f{1}{2} \left( \p_\lambda \beta_\nu - \p_\nu \beta_\lambda \right).
\label{varpi1}
\eea
Hence, \EQSTWO{leqBel}{leqcan} are equivalent only if $\omega_{\lambda\nu} = \varpi_{\lambda\nu} $.  This fact indicates that the description based on the Belinfante tensors is reduced compared to the description employing the canonical tensors. 
 
\subsection{Global-equilibrium density operator}
\label{sect:geqop}
\medskip

In global equilibrium the integral in the argument of the exponential function in \EQ{leqcan} should be independent of the choice of the space-like hypersurface $\Sigma$. To check when this happens, we rewrite \EQ{leqcan} introducing the total angular momentum operator. In this way we obtain
\bea
\hspace{-0.5cm} \rhoLEQ = \f{1}{Z} \exp \left[-\int_\Sigma \di \Sigma_\mu 
\left(\oTmnUc  \left(\beta_\nu - \omega_{\nu\lambda} x^\lambda \right)
- \f{1}{2} \omega_{\lambda\nu}  \oJmlnUc  - \xi \wj^\mu  \right) \right].
\label{leqcan2}
\eea
The operator $\rhoLEQ$ is constant if the divergence of the integrand in \EQ{leqcan2} vanishes. Since the energy-momentum and angular-momentum tensors are conserved, $\p_\mu \oTmnUc  =0$ and $\p_\mu  \oJmlnUc= 0$, we obtain three conditions for the Lagrange multipliers~\footnote{We use here the fact that $\oTmnUc$ is not symmetric and $\oJmlnUc$ has no additional properties than asymmetry in the last two indices.}:
\bea
\p_\mu \omega_{\nu\lambda} = 0, \qquad \p_\mu \beta_\nu = \onmL , \qquad \p_\mu \xi = 0,
\label{geqcond1}
\eea
which imply that $\xi$ and the spin polarization tensor $\onmL$ should be constant, whereas the $\beta_\nu$ field should have the form
\bea
\beta_\nu = b^0_\nu + \omega_{\nu \lambda} x^\lambda.  \label{geqcond2}
\eea
Equations \EQn{geqcond1} and \EQn{geqcond2} imply that the $\beta_\nu$ field  satisfies the Killing equation
\bea
\p_\mu \beta_\nu + \p_\nu \beta_\mu = 0 
\label{killing1}
\eea
and the spin polarization tensor is equal to thermal vorticity, $\omega_{\lambda\nu} = \varpi_{\lambda\nu}$. Thus, the global-equilibrium statistical operator has the form
\bea
\rhoGEQ = \f{1}{Z} \exp \left[-\int_\Sigma \di \Sigma_\mu \left(\oTmnUc  \beta_\nu 
- \f{1}{2} \varpi_{\lambda\nu} \oSmlnUc  - \xi \wj^\mu\right) \right],  \label{geqcan}
\eea
where $ \varpi_{\lambda\nu} = -\f{1}{2} (\p_\lambda \beta_\nu-\p_\nu \beta_\lambda) = \hbox{const}$.

\subsection{General concept of perfect-fluid hydrodynamics with spin}
\label{sect:pfspin}
\medskip

If the conditions \EQn{geqcond1} are not satisfied, the integral over the hypersurface $\Sigma$ depends on its choice. The two such integrals, over the hypersurfaces $\Sigma_1$ and $\Sigma_2$ differ by the volume integral. One can show, however, that such a volume integral describes dissipative phenomena~\CITn{Zubarev:1966aa,Zubarev:1979aa}, hence, if we neglect dissipation we may treat the local-equilibrium operator \EQn{leqcan} as constant. In this case we define the expectation values of the conserved currents through the expressions:
\bea
\hspace{-0.5cm} 
\TmnU =  \tr \left(\rhoLEQ \, \oTmnU \right), \quad
\SmlnU = \tr \left(\rhoLEQ \,  \oSmlnU \right), \quad
j^{\mu} = \tr \left(\rhoLEQ \, \wj^{\mu}\right).
\label{TSj1}
\eea
They are all functions of the hydrodynamic variables $\beta_\mu$, $\omega_{\mu\nu}$, and $\xi$, namely,
\bea
\hspace{-0.5cm} 
\TmnU = \TmnU [\beta,\omega,\xi], \quad
\SmlnU = \SmlnU [\beta,\omega,\xi], \quad
j^\mu = j^\mu [\beta,\omega,\xi],
\label{TSj2}
\eea
and satisfy the conservation laws
\bea
\p_\mu \TmnU = 0, \quad
\p_\lambda \SlmnU = \TnmU -\TmnU, \quad
\p_\mu  j^\mu =0.
\label{TSj3}
\eea
In general, these are 11 equations for 11 unknown functions, which represent a generalization of the standard perfect-fluid hydrodynamics to the case including spin. We also stress that below we shall deal with the expectation values defined by \EQS{TSj1} rather than with operators.

\section{Local equilibrium functions \label{chapt:leqfun}}

\subsection{Spin dependent phase-space distribution functions}
\label{sec:spinphasespace}
\medskip

In this section we put forward the phase-space distribution functions for spin-$\onehalf$ particles and antiparticles in local equilibrium, which have been introduced by Becattini et~al. in  \CIT{Becattini:2013fla}. In this approach, to include spin degrees of freedom, the standard scalar functions are generalized to 2$\times$2 Hermitean matrices in spin space for each value of the space-time position $x$ and four-momentum $p$,
\bea
\left[ f^+(x,p) \right]_{rs}  \equiv  \fplusrsxp &=&  \ubarrp X^+ \usp, \label{fplusrsxp}  \\
\left[ f^-(x,p) \right]_{rs}  \equiv \fminusrsxp &=& - \vbarsp X^- \vrp. \label{fminusrsxp}
\eea
Here $\urp$ and $\vrp$ are Dirac bispinors (with the spin indices $r$ and $s$ running from 1~to~2), and the normalization $\ubarrp \usp=\,\delta_{rs}$ and $\vbarrp \vsp=-\,\delta_{rs}$. Note the minus sign and different ordering of spin indices in \EQ{fminusrsxp} compared to \EQ{fplusrsxp}. 

The 4$\times$4 matrices $X^{\pm}$ in \EQSTWO{fplusrsxp}{fminusrsxp} are defined as products of the relativistic Boltzmann distributions (J\"uttner distributions~\CITn{Juttner:1911aa}) and matrices $M^\pm$, namely
\bea
X^{\pm} =  \exp\left[\pm \xi(x) - \bmL(x) \pmU \right] M^\pm,  \label{XpmM}
\eea
where
\bea
M^\pm = \exp\left[ \pm \f{1}{2} \omnL(x)  \SmnU \right] .   \label{Mpm}
\eea
In \EQSTWO{XpmM}{Mpm} $\beta^\mu$ denotes, as above, the ratio between the fluid four-velocity $\umU$ and the local temperature $T$, while $\xi$ is the ratio between the chemical potential $\mu$ and $T$. The quantity $\omnL$ is the spin  polarization tensor introduced in the previous section. For the sake of simplicity, we restrict ourselves herein to classical Boltzmann statistics. The quantum statistics may be included by replacing the exponential functions in \EQSTWO{XpmM}{Mpm} by the Fermi-Dirac distribution. This approach was proposed and worked out in  \CIT{Becattini:2013fla}.

At this point it should be recalled that Eq.~(\ref{Mpm}) was originally introduced for the case where the spin polarization tensor $\omega_{\mu\nu}$ is equal to the thermal vorticity $\varpi_{\mu\nu}$. In this work, following the arguments given in Sec.~\ref{sect:pfspin}, we relax this condition and allow $\omega_{\mu\nu}$ to be treated as an independent hydrodynamic variable (strictly speaking, as a set of six independent hydrodynamic variables). Our results presented below (based on different formulations of kinetic theory) show that it is a reasonable assumption as long as we restrict ourselves to a perfect-fluid picture.  

In analogy to the Faraday tensor $F_{\mu\nu}$ used in classical electrodynamics, the antisymmetric spin polarization tensor $\omnL$ can be always defined  in terms of electric- and magnetic-like three-vectors in LAB, $\ev = (e^1,e^2,e^3)$ and $\bv = (b^1,b^2,b^3)$. In this case we write (following the electrodynamic sign conventions of \CIT{Jackson:1998nia})
\bel{omeb}
\omnL= 
\begin{bmatrix}
	0       &  e^1 & e^2 & e^3 \\
	-e^1  &  0    & -b^3 & b^2 \\
	-e^2  &  b^3 & 0 & -b^1 \\
	-e^3  & -b^2 & b^1 & 0
\end{bmatrix}.
\label{eq:omegaeb}
\eel
To switch from $\omnL$ to the dual tensor $\omnLD$, one replaces $\ev$ by $\bv$ and $\bv$ by~$-\ev$. Hence, the parametrization of the dual spin polarization tensor reads
\bel{omdeb}
\omnLD = 
\begin{bmatrix}
	0       &  b^1 & b^2 & b^3 \\
	-b^1  &  0    & e^3 & -e^2 \\
	-b^2  &  -e^3 & 0 & e^1 \\
	-b^3  & e^2 & -e^1 & 0
\end{bmatrix}.
\label{eq:omegaebd}
\eel
One also finds that
\bea
\f{1}{2} \omnL \omnU = -\f{1}{2} \omnLD \omnUD 
=   \bv \cdot \bv - \ev \cdot \ev, \quad
\omnLD \omnU = -4 \ev \cdot \bv . \label{om3} 
\eea

Before we continue our discussion, it is important to stress that the forms of the equilibrium distributions with spin given by \EQSTWO{fplusrsxp}{fminusrsxp} have not been derived yet from any underlying microscopic model or theory. In particular, it is not known at the moment if they can be obtained from a kinetic-theory approach. To large extent, \EQSTWO{fplusrsxp}{fminusrsxp} represent an ``educated guess'' for the form of equilibrium phase-space distributions with spin-$\onehalf$. In the present work, by discussing the quantum kinetic equation in the semi-classical expansion and by developing the hydrodynamic framework with a classical treatment of spin, we eventually conclude on the applicability range of \EQSTWO{fplusrsxp}{fminusrsxp}. Nonetheless, before concluding on this issue, we develop a formalism based on \EQSTWO{fplusrsxp}{fminusrsxp}, as this allows us to build a necessary theoretical scheme and to introduce relevant definitions and concepts.

\subsection{Important special cases}
\label{sec:specM}
\medskip

The exponential dependence of the distribution function on the Dirac spin operator $\Sigma^{\mu\nu}$ given in \EQ{Mpm}, which is defined in terms of a power series, can be resummed. This results in the following expression for $M^\pm$~\CITn{Florkowski:2017dyn}, 
\bea
M^{\pm} &=&  \boldsymbol{1} \left[ \Re(\cosh z) \pm \Re \LB \frac{\sinh z}{2 z} \RB \omega_{\mu \nu}  \Sigma^{\mu \nu}\right]
\nonumber \\
&&\qquad\qquad+ i \gamma_5 \left[ \Im(\cosh z)\pm \Im \LB \frac{\sinh z}{2 z} \RB \omega_{\mu \nu}   \Sigma^{\mu \nu} \right],
\label{MpmExp}
\eea
where $\boldsymbol{1}$ is a unit 4$\times$4 matrix and
\bea
z  &=&  \f{1}{2 \sqrt{2}} \sqrt{ \omnL \omnU + i \omnL \omnUD} \,.
\label{z}
\eea

In a series of papers~\CITn{Florkowski:2017ruc,Florkowski:2017dyn,Florkowski:2018ual}, the special case was analyzed where elements of the spin polarization tensor fulfill the conditions~\footnote{The conditions \EQn{TRcase} have been relaxed in a series of recent papers, for example, see \CITS{Florkowski:2018myy,Prokhorov:2018qhq,Prokhorov:2018bql}.}
\bea
\ev \cdot \bv = 0, \qquad  \bv \cdot \bv - \ev \cdot \ev \geq 0. \label{TRcase}
\eea
If \EQS{TRcase} are satisfied, the variable $z$ becomes a real number $\zeta$ and \EQ{MpmExp} simplifies to the expression linear in the operator $ \SmnU$, namely
\bea
M^\pm &=& \cosh(\zeta) \pm  \f{\sinh(\zeta)}{2\zeta}  \, \omnL \SmnU  , \label{MpmTR}
\eea
where
\bea
\zeta =  \f{1}{2 \sqrt{2}} \sqrt{ \omnL \omnU}.
\label{zeta}
\eea
Another interesting case corresponds to small polarization. If  $\omnL \ll 1$ we can use the expression
\bea
M^\pm &=& 1 \pm  \f{1}{2}  \, \omnL \SmnU, \label{Mpm0}
\eea
where the elements of the polarization tensor are not constrained by the conditions~\EQn{TRcase}. One can easily notice that \EQ{Mpm0} follows directly from the expansion of the exponential function \EQn{Mpm} up to the first order in $\omnL$ and does not constrain otherwise the coefficients $\omnL$. Below, most of the calculations are done with $M^\pm$ defined by \EQ{MpmTR} as it reproduces \EQ{Mpm0} as a special case.

\subsection{Spin polarization three-vector}
\label{sec:P3V}
\medskip

The spin observables are represented by the Pauli matrices $\sigv$ and the expectation values of $\sigv$ provide information on the polarization of spin-$\onehalf$ particles in their rest frame. Since we consider Dirac bispinors obtained by the canonical Lorentz boosts applied to states with zero three-momentum, we refer to the resulting spin distributions and particle rest frames as the canonical ones (they differ from other definitions by a rotation).

Using \EQ{MpmTR}, the spin dependent distribution functions defined by \EQSTWO{fplusrsxp}{fminusrsxp} can be rewritten in a form linear in the Dirac spin tensor,
\bel{fplusrsxp1}
\fplusrsxp =  e^{\xi - p \cdot \beta} \left[ \cosh(\zeta) \delta_{rs} + \f{\sinh(\zeta)}{2\zeta}  \, \ubarrp \oabL \SabU \usp \right],
\eel
\bel{fminusrsxp1}
\fminusrsxp = e^{-\xi - p \cdot \beta} \left[ \cosh(\zeta) \delta_{rs} + \f{\sinh(\zeta)}{2\zeta}  \,  \vbarsp \oabL \SabU \vrp \right].
\eel
We then find a compact expression~\CITn{Florkowski:2017dyn}
\bel{fpm}
f^\pm(x,p) =  e^{\pm \xi - p \cdot \beta} \left[ \cosh(\zeta)  - \f{\sinh(\zeta)}{2\zeta}  \, \Pv \cdot \sigv \right],
\eel
where we have introduced a polarization vector 
\bea
\Pv &=& \f{1}{m} \left[  E_p \, \bv - \pv \times \ev - \f{\pv \cdot \bv}{E_p + m} \pv \right].
\label{Aveb}
\eea
The three-vector $\Pv$ can be interpreted as a spatial part of the polarization four-vector $P^\mu$, with a vanishing zeroth component, $P^\mu(x,p)=\left(0,\Pv(x,p)\right)$. The average polarization vector  is defined by the formula
\bea
\left\langle \Pv(x,p) \right\rangle =  \f{1}{2} \f{ \trt \left[ (f^+ + f^-) \sigv\right]  }{\trt \left[ f^+ + f^- \right] }  = -\f{1}{4 \zeta} \tanh(\zeta) \Pv .
\label{eq:avPv1}
\eea
We note that the expression on the right-hand side of \EQ{Aveb} defines the field $\bv$ in the particle rest frame~\CITn{Jackson:1998nia}. 
We summarize this finding by writing
\bea 
\Pv &=&  \bv_\ast .
\label{AvebPRF}
\eea
Thus, the polarization is determined by the value of the field $\bv$ in the particle canonical rest frame. Moreover, using \EQ{zeta}, we may obtain an alternative expression for the average polarization vector
\bea
\left\langle \Pv(x,p) \right\rangle = - \f{1}{2} \tanh\left[  \f{1}{2} \sqrt{ \bv_\ast \cdot \bv_\ast - \ev_\ast \cdot \ev_\ast } \right]  
\f{ \bv_\ast }{\sqrt{ \bv_\ast \cdot \bv_\ast - \ev_\ast \cdot \ev_\ast }}  ,
\label{eq:avPv2}
\eea
where we have used the fact that the quantity $\bv \cdot \bv - \ev \cdot \ev$ is independent of the choice of the Lorentz frame. 

Equation~\EQn{eq:avPv2} indicates a potential problem with the application of the present formalism for particles with the momentum $p$, which are placed at the space-time point $x$ and for which the condition
\bea 
|\left\langle \Pv(x,p) \right\rangle| \leq \f{1}{2}
\label{eq:Pvcond}
\eea
is violated. The condition \EQn{eq:Pvcond} holds, for example, if for all particles under consideration the magnetic-like components dominate $|\ev_\ast| <  |\bv_\ast|$ or if the spin polarization components are sufficiently small, $\zeta < 1$, and $|\bv_\ast| <  1$~\footnote{Strictly speaking, there are always particles in a system, which have large momenta and violate the condition $|\bv_\ast| < 1$. However, their abundances are usually negligible, since suppressed by a thermal factor. The only exception are particles that move fast together with a fluid element. In this case it is reasonable to replace the condition $|\bv_\ast| \ll 1$ by a more precise requirement that $|\bv_\ast| \gamma < 1$, where $\gamma$ is the Lorentz factor connected with a fluid velocity. }. Such constraints can be always checked for systems under investigations. We shall come back to a discussion of the condition \EQn{eq:Pvcond} in Sec.~\ref{sec:class-PL}, where we reexamine it in the context of classical description of spin.

\section{Semi-classical expansion of Wigner function} 
\label{sec:expWig}

In this section we study consequences of using \EQSTWO{fplusrsxp}{fminusrsxp} as an input for construction of the equilibrium Wigner function. Our approach closely follows recent investigations presented in \CIT{Florkowski:2018ahw} --- we first recapitulate the steps leading to the kinetic equations satisfied by the coefficients of the equilibrium Wigner function (in the Clifford-algebra representation) and, subsequently, argue that simple moments of the kinetic equations lead to the GLW hydrodynamic picture. 

\subsection{Equilibrium Wigner functions} 
\label{sec:eqWig}
\medskip

The functions $f^\pm_{rs}(x,p)$ can be used to determine explicit expressions for the corresponding equilibrium (particle and antiparticle) Wigner functions ${\cal W}^\pm_{\rm eq}(x,k)$. We construct them using the relations derived in \CIT{DeGroot:1980dk},
\bea
\Weqpxk = \frac{1}{2} \sum_{r,s=1}^2 \int dP\,
\delta^{(4)}(k-p) u^r(p) {\bar u}^s(p) f^+_{rs}(x,p),
\label{eq:Weqpxk}
\eea
\bea
\Weqmxk = -\frac{1}{2} \sum_{r,s=1}^2 \int dP\,
\delta^{(4)}(k+p) v^s(p) {\bar v}^r(p) f^-_{rs}(x,p).
\label{eq:Weqmxk}
\eea
Here $k$ is the four-momentum of particles or antiparticles, and $dP$ is the invariant integration measure defined by~\EQ{eq:dP}. The total Wigner function is a simple sum of the particle and antiparticle contributions
\bel{eq:totW}
\Weqxk = \Weqpxk + \Weqmxk.
\eel
Using \EQSTWO{fplusrsxp}{fminusrsxp} we find
\bea
{\cal W}^\pm_{\rm eq}(x,k) = \frac{1}{4 m}  \int dP\,
\delta^{(4)}(k \mp p) (\slashed{p} \pm m) X^\pm (\slashed{p} \pm m).
\label{eq:Weqpxk1}
\eea
With the help of \EQ{MpmTR} we can further rewrite this equation as
\bea
{\cal W}^\pm_{\rm eq}(x,k) &=& \frac{e^{\pm \xi}}{4 m}  \int dP
\,e^{-\beta \cdot p }\,\, \delta^{(4)}(k \mp p) \label{eq:Weqpxk2} \\
&& \times  \left[2m (m \pm \slashed{p}) \cosh(\zeta) \pm \f{\sinh(\zeta)}{2\zeta}  \, \omnL \,(\slashed{p} \pm m) \SmnU (\slashed{p} \pm m) \right]. \nn
\eea

The presence of the Dirac delta functions in the definitions of the equilibrium Wigner functions \EQn{eq:Weqpxk2} indicates that, to large extent, they describe classical motion --- the energy of particles is always on the mass shell. This suggests that the functions \EQn{eq:Weqpxk2} cannot be regarded as complete, quantum-mechanical equilibrium distributions. In fact, we shall see below that the expressions \EQn{eq:Weqpxk2} can be identified only with the leading order terms in $\hbar$ of the ``true'' equilibrium Wigner functions which satisfy the quantum kinetic equation. For our approach it is important, however, that the functions \EQn{eq:Weqpxk2} incorporate spin degrees of freedom and may serve to construct the formalism of hydrodynamics with spin (in the leading order in $\hbar$). 

\subsection{Clifford-algebra expansion} 
\label{sec:clifford}
\medskip

The Wigner functions $\Wpmxk$ are 4$\times$4 matrices satisfying the conjugation relation $\Wpmxk = \gamma_0 \Wpmxk^\dagger \gamma_0$. Consequently, they can be expanded in terms of 16 independent generators of the Clifford algebra  with real coefficients~\cite{Itzykson:1980rh}. Such an expansion method was used very successfully in the past to formulate the transport equations for abelian plasmas~\CITn{Vasak:1987um,Zhuang:1995pd}, the quark-gluon plasma~\CITn{Elze:1986hq,Elze:1986qd}, and chiral models~\CITn{Florkowski:1995ei}. More recently, it has been used, for example,  in~\CITS{Gao:2012ix,Fang:2016vpj,Fang:2016uds,Huang:2018wdl}. 

In our case,  we start with the decomposition of the equilibrium Wigner function \EQn{eq:Weqpxk2} in the form
\bea
\Weqpmxk &=& \f{1}{4} \left[ \Feqpmxk + i \gamma_5 \Peqpmxk + \gamma^\mu {\cal V}^\pm_{{\rm eq}, \mu}(x,k) \right. \nn \\
&& \left.  \hspace{1cm} + \gamma_5 \gamma^\mu {\cal A}^\pm_{{\rm eq}, \mu}(x,k)
+ \SmnU {\cal S}^\pm_{{\rm eq}, \mu \nu}(x,k) \right].
\label{eq:wig_expansion}
\eea
The coefficient functions in the expansion~\EQn{eq:wig_expansion} can be obtained in the straightforward way by calculating the trace of $\Weqpmxk$ multiplied first by the matrices: $\boldsymbol{1}$, $-i \gfive, \gamma_\mu, \gamma_\mu \gfive$ and $2\,\SmnL$. In this way we find the expressions~\CITn{Florkowski:2018ahw}:
\bea
\Feqpmxk &=& 2 m \cosh(\zeta)\,\int dP\, \,e^{-\beta \cdot p \pm \xi}\,\,\delta^{(4)} (k\mp p),  \label{eq:FEeqpm} \\
\Peqpmxk &=& 0, \label{eq:PEeqpm} \\
{\cal V}^{\pm}_{{\rm eq}, \mu}(x,k) &=& \pm\,2 \cosh(\zeta) \, \int dP\,e^{-\beta \cdot p \pm \xi}\,\,\delta^{(4)} (k\mp p)\,p_{\mu}, \label{eq:VEeqpm} \\
{\cal A}^\pm_{{\rm eq}, \mu}(x,k) &=& -\frac{\sinh(\zeta)\, }{\zeta} \,\int dP\,e^{-\beta \cdot p \pm \xi}\,\,\delta^{(4)}(k\mp p)\, \tilde{\omega }_{\mu \nu}\,p^{\nu},
\label{eq:AEeqpm} \\
{\cal S}^\pm_{{\rm eq}, \mu \nu}(x,k) &=& \! \pm\frac{ \sinh(\zeta) }{m \zeta} \!\int \!dP\,e^{-\beta \cdot p \pm\xi}\,\,\delta^{(4)}(k\mp p) \label{eq:SEeqpm}  \\
&& \hspace{2cm} \times \left[  \left( p_\mu \omega_{\nu \alpha} -  p_\nu \omega_{\mu \alpha} \right) p^\alpha \!+\! m^2\omega_{\mu \nu} \right]. \nn
\eea
One can easily check that the functions defined by~\EQSM{eq:FEeqpm}{eq:SEeqpm} satisfy the following set of constraints:
\bel{eq:Wid12}
k^\mu \, {\cal V}^{\pm}_{{\rm eq}, \mu}(x,k) = 
m \, {\cal F}^{\pm}_{{\rm eq}}(x,k), \quad 
k_\mu \, {\cal F}^{\pm}_{{\rm eq}}(x,k) = m \, {\cal V}^{\pm}_{{\rm eq}, \mu}(x,k),
\eel
\bel{eq:Wid345}
{\cal P}^\pm_{{\rm eq}}(x,k) = 0, \quad k^\mu \, {\cal A}^{\pm}_{{\rm eq}, \,\mu}(x,k) = 0, \quad k^\mu \, {\cal S}^{\pm}_{{\rm eq}, \,\mu \nu}(x,k) = 0,
\eel
\bel{eq:Wid6}
k^\beta \, {\tilde {\cal S}}^{\pm}_{{\rm eq}, \mu \beta}(x,k) + m \, {\cal A}^{\pm}_{{\rm eq}, \,\mu}(x,k) = 0,
\eel
\bel{eq:Wid7}
\epsilon_{\mu \nu \alpha \beta} \, k^\alpha \, {\cal A}^{\pm \, \beta}_{{\rm eq}}(x,k) + m \, {\cal S}^{\pm}_{{\rm eq}, \,\mu \nu}(x,k) = 0.
\eel
We note that such constraints are fulfilled also by the total Wigner function given by the sum of particle and antiparticle contributions, see \EQ{eq:totW}. We also note that \EQSM{eq:Wid12}{eq:Wid7} follow from the algebraic structure of the equilibrium Wigner functions and are satsified for any form of the fields: $\beta_\mu(x)$, $\xi(x)$, and $\omega_{\mu \nu}(x)$. 

\subsection{Global equilibrium}
\label{sect:geqkin}
\medskip

Generally speaking, the Wigner function ${\cal W}(x,k)$ satisfies the kinetic equation that can be schematically written as
\bel{eq:eqforWC}
\left(\gamma_\mu K^\mu - m \right) {\cal W}(x,k) = C[{\cal W}(x,k)].
\eel
Here $K^\mu$ is the operator defined by the expression
\bel{eq:K}
K^\mu = k^\mu + \frac{i \hbar}{2} \,\p^\mu,
\eel
whereas $C[{\cal W}(x,k)]$ is the collision term. We tacitly assume that $C[{\cal W}(x,k)]$ vanishes if ${\cal W}(x,k)$ describes any form of equilibrium, global or local. 

In the case of global equilibrium, with the vanishing collision term, the Wigner function ${\cal W}(x,k)$ exactly satisfies the equation~\footnote{Here we neglect the presence of the electromagnetic and other mean fields. Their inclusion is straightforward and left for future studies generalizing the present framework.}
\bel{eq:eqforW}
\left(\gamma_\mu K^\mu - m \right) {\cal W}(x,k) = 0.
\eel
The standard treatment of this equation is based on the semi-classical expansion in powers of $\hbar$ of the coefficient functions defining ${\cal W}(x,k)$. Such an expansion shows that one can choose, as two independent functions, the coefficients $ {\cal F}_{(0)}(x,k)$ and ${\cal A}^\nu_{(0)} (x,k)$ --- the other coefficient functions are defined in terms of these two functions only. Moreover, one can check that algebraic relations imposed by \EQ{eq:eqforW} in the leading order (LO) in $\hbar$ are satisfied by the equilibrium functions  \EQSMn{eq:FEeqpm}{eq:SEeqpm}. Hence, one can assume that:
\bea
{\cal F}^{(0)} &=& {\cal F}_{\rm eq},
\label{eq:FC1} \\
{\cal P}^{(0)} &=&0,
\label{eq:PC1} \\
{\cal V}^{(0)}_\mu  &=& {\cal V}_{\rm eq, \mu},
\label{eq:VC1}\\
{\cal A}^{(0)}_\mu  &=& {\cal A}_{\rm eq, \mu},
\label{eq:AC1}\\
{\cal S}^{(0)}_{\mu \nu} &=& {\cal S}_{{\rm eq}, \mu \nu}.
\label{eq:SC1} 
\eea
The next-to-leading-order (NLO) terms in $\hbar$ coming from \EQ{eq:eqforW} yield the dynamic equations
\bel{eq:kineqFC1}
k^\mu \p_\mu {\cal F}_{\rm eq}(x,k) = 0,
\eel
\bel{eq:kineqAC1}
k^\mu \p_\mu \, {\cal A}^\nu_{\rm eq} (x,k) = 0, 
\quad k_\nu \,{\cal A}^\nu_{\rm eq}(x,k) = 0,
\eel
which are nothing else but the kinetic equations that should be obeyed by the coefficient functions ${\cal F}_{\rm eq}(x,k)$ and $ {\cal A}^\nu_{\rm eq} (x,k)$. Since the functions ${\cal F}_{\rm eq}(x,k)$ and ${\cal A}^\nu_{\rm eq}(x,k)$ depend separately on $p^\alpha \beta_\alpha(x)$, $\xi(x)$ and $\omega_{\mu\nu}(x)$, one can easily check that \EQSTWO{eq:kineqFC1}{eq:kineqAC1} are satisfied if $\beta_\mu$ is a Killing vector, see~\EQ{killing1}, while $\xi$ and $\omega_{\mu\nu}$ are constant.  However, the thermal vorticity obtained with  the field $\beta_\mu$ is not necessarily equal to $\omega_{\mu\nu}$, although the two tensors should be constant. We also note that if Eqs.~(\ref{eq:kineqFC1}) and (\ref{eq:kineqAC1}) are satisfied, then the Wigner function (\ref{eq:totW}), constructed from ${\cal F}_{\rm eq}(x,k)$ and $ {\cal A}^\nu_{\rm eq} (x,k)$, satisfies the kinetic equation (\ref{eq:eqforWC}).

Our general arguments presented earlier, see Sec.~\ref{sect:geqop}, indicated that thermal vorticity should be equal to the spin polarization tensor in global thermodynamic equilibrium, so how these two facts can be reconciled? The most likely answer to this question is that the equality of the spin polarization tensor and thermal vorticity is a result of dissipative phenomena that are neglected in the present approach~\footnote{Note also that the equality between the spin polarization tensor and thermal vorticity requires an asymmetric energy-momentum tensor. This issue will be discussed below.}.

In view of the present discussion we may distinguish between global and extended global equilibria~\CITn{Florkowski:2018ahw}: In global equilibrium the $\beta_\mu$ field is a Killing vector satisfying \EQ{killing1}, $\varpi_{\mu \nu} = -\frac{1}{2} \left(\p_\mu \beta_\nu - \p_\nu \beta_\mu \right) = \hbox{const}$, the spin polarization tensor is constant and agrees with thermal vorticity, $\omega_{\mu\nu} = \varpi_{\mu \nu}$, in addition $\xi = \hbox{const}$. In an extended global equilibrium $\beta_\mu$ field is a Killing vector, $\varpi_{\mu \nu} = -\frac{1}{2} \left(\p_\mu \beta_\nu - \p_\nu \beta_\mu \right) = \hbox{const}$, the spin polarization tensor is constant but $\omega_{\mu\nu} \neq \varpi_{\mu \nu}$, $\xi = \hbox{const}$. Correspondingly, we also differentiate between local and extended local equilibria. In local equilibrium the $\beta_\mu$ field is not a Killing vector but we still have $\omega_{\mu\nu}(x) = \varpi_{\mu \nu}(x)$, $\xi$ is allowed to depend on space-time coordinates, $\xi = \xi(x)$. In extended local equilibrium the $\beta_\mu$ field is not a Killing vector and $\omega_{\mu\nu}(x) \neq \varpi_{\mu \nu}(x)$, moreover $\xi = \xi(x)$. We note that \EQSM{eq:Wid12}{eq:Wid7} follow from the algebraic structure of the equilibrium Wigner functions and are satsified for any form of the fields: $\beta_\mu(x)$, $\xi(x)$, and $\omega_{\mu \nu}(x)$. Thus, they hold for four different types of equilibrium specified above.

\subsection{Perfect-fluid hydrodynamics with spin from kinetic theory}
\label{sect:pfspinkin}
\medskip

Equations of the perfect-fluid hydrodynamics with spin can be obtained by approximate treatment of \EQ{eq:kineqFC1} and \EQ{eq:kineqAC1}. In this case we do not require that they are  exactly satisfied but demand instead that certain moments of them (in the momentum space) vanish, allowing for space-time dependence of the hydrodynamic variables $\beta_\mu$, $\xi$, and $\omnL$. 

Dealing with \EQ{eq:kineqFC1} is quite well established --- one has to include its zeroth and first moments, which leads to the conservation of charge, energy, and linear momentum. Indeed, the four-dimensional integration of \EQ{eq:kineqFC1} over $k$ yields
\bel{eq:Ncon}
\p_\alpha N^\alpha_{\rm eq}(x)  = 0,
\eel
where
\bel{eq:Nalpha1}
N^\alpha_{\rm eq} = 4\cosh(\zeta) \sinh(\xi)  \int \frac{d^3p}{(2\pi)^3 E_p} \, p^\alpha
\,e^{-\beta \cdot p }.
\eel
Doing the integral over the three-momentum in \EQ{eq:Nalpha1}, one finds that the charge current is proportional to the flow vector,
\bel{Nmu}
N^\alpha_{\rm eq} = n u^\alpha,
\eel
where 
\bel{nden}
n = 4 \, \cosh(\zeta) \sinh(\xi)\, \n0(T)
\eel
is the charge density~\CITn{Florkowski:2017ruc}. The quantity 
\bel{nden0}
\n0(T) = \langle(u\cdot p)\rangle_0
\eel
is the number density of spin-0, neutral Boltzmann particles, obtained using the thermal average
\bel{avdef}
\langle \cdots \rangle_0 \equiv \int \f{d^3p}{(2\pi)^3 E_p}  (\cdots) \,  e^{- \beta \cdot p}.
\eel
If we multiply \EQ{eq:kineqFC1} first by $k^\nu$ and then perform the four-dimensional integration over $k$, we obtain the conservation of energy and momentum in the form
\bel{eq:Tcon}
\p_\alpha T^{\alpha\beta}_{\rm eq}(x) = 0,
\eel
where the energy-momentum tensor is defined by the perfect-fluid formula
\bel{Tmn}
T^{\a\b}_{\rm eq}(x) &=& (\varepsilon + P ) u^\a u^\b - P g^{\a\b},
\eel
with 
\bel{enden}
\varepsilon = 4 \, \cosh(\zeta) \cosh(\xi) \, \e0(T)
\eel
and
\bel{prs}
P = 4 \, \cosh(\zeta) \cosh(\xi) \, \P0(T),
\eel
respectively~\CITn{Florkowski:2017ruc}. In analogy to the density $\n0(T)$, we define the auxiliary 
quantities
\bel{enden0}
\e0(T) &=& \langle(u\cdot p)^2\rangle_0
\eel
and
\bel{prs0}
\P0(T) = -(1/3) \langle \left[ p\cdot p - (u\cdot p)^2 \right] \rangle_0. 
\eel

The construction of the hydrodynamic-like equation from the conservation of the axial current \EQn{eq:kineqAC1} is less obvious. We follow here the treatment of ~\CIT{Florkowski:2018ahw} and first rewrite \EQ{eq:kineqAC1} as
\bea
0 &=& k^\alpha  \p_\alpha  \, 
\,\int dP\,e^{-\beta \cdot p }\, \frac{\sinh(\zeta)\, }{\zeta} 
\left[ \delta^{(4)}(k-p) e^{\xi}
+ \delta^{(4)}(k+p) e^{-\xi} \right]
\, \tilde{\omega }_{\mu \nu}\,p^{\nu} .
\label{eq:h0} 
\eea
In the next step we multiply \EQ{eq:h0} by the four-vector $k_\beta$, contract it with the Levi-Civita tensor $\epsilon^{\mu\beta\gamma\delta}$ and, finally, integrate the resulting equation over $k$. In this way we obtain the conservation law of the form,
\bea
\p_\lambda S^{\lambda , \mu \nu }_{\rm eq}(x) = 0,
\label{eq:SGLWcon}
\eea
with
\bea
S^{\lambda , \mu \nu }_{\rm eq}&=&\frac{\hbar \sinh (\zeta) {\cosh}(\xi)}{m^2\zeta }\int dP \, e^{-\beta \cdot p} p^{\lambda } \left(m^2\omega ^{\mu\nu}+2 p^{\alpha }p^{[\mu }\omega ^{\nu ]}{}_{\alpha } 
\right)  \label{eq:Smunulambda_de_Groot22} \nn  \\
 &=&  \frac{\hbar w}{4 \zeta} u^\lambda \omega^{\mu\nu}  +  {\cal C} S^{\lambda , \mu \nu }_{\Delta} 
 = {\cal C} \left( n_{(0)}(T) u^\lambda \omega^{\mu\nu}  +  S^{\lambda , \mu \nu }_{\Delta} \right),
\label{eq:Smunulambda_de_Groot2}
\eea 
where ${\cal C}= \hbar  \ch{\xi}\sh{\zeta}/\zeta$. In the last line we have used the spin density $w$ defined in~\cite{Florkowski:2017ruc}
\bel{eq:w}
w = 4 \sinh(\zeta) \cosh(\xi) n_{(0)}(T)
\eel
and the auxiliary tensor
\beq
S^{\a, \b\g}_{\Delta} 
&=& \label{SDeltaGLW} {\cal A}_{(0)} \, u^\a u^\d u^{[\b} \omega^{\g]}_{\HP\d}    + \, {\cal B}_{(0)} \, \Big( 
u^{[\b} \Delta^{\a\d} \omega^{\g]}_{\HP\d}
+ u^\a \Delta^{\d[\b} \omega^{\g]}_{\HP\d}
+ u^\d \Delta^{\a[\b} \omega^{\g]}_{\HP\d}\Big),
\eeq
where
\beq 
{\cal B}_{(0)} &=&-\frac{2}{\hat{m}^2}  \frac{\varepsilon_{(0)}(T)+P_{(0)}(T)}{T}=-\frac{2}{\hat{m}^2} s_{(0)}(T)\label{coefB}
\eeq
and
\beq
{\cal A}_{(0)} &=&\frac{6}{\hat{m}^2} s_{(0)}(T) +2 n_{(0)} (T) = -3{\cal B}_{(0)} +2 n_{(0)}(T),
\label{coefA}
\eeq
with $\sU =   \LR\eU+\PU\RR / T$ being the entropy density. The operator projecting on the space orthogonal to the flow four-vector  is defined by $\Delta^{\mu\nu} =g^{\mu\nu} - u^\mu u^\nu$.  It is important to note that for dimensional reasons, we have implemented the Planck constant $\hbar$ in the definition \EQn{eq:Smunulambda_de_Groot2}.

We summarize this section with the observation that the use of the equilibrium Wigner functions (\ref{eq:Weqpxk}) and (\ref{eq:Weqmxk}) in the kinetic equation (\ref{eq:eqforWC}) allowed us to construct the two conservation laws, Eqs.~(\ref{eq:Tcon}) and (\ref{eq:SGLWcon}), for the tensors $T^{\mu\nu}_{\rm eq}(x) $ and $S^{\lambda , \mu \nu }_{\rm eq}$.
In the next section, we shall find that the forms of these tensors agree with the expressions provided in~\CIT{DeGroot:1980dk}.

\section{Conserved currents \label{chapt:concur}}

\subsection{NLO corrections to Wigner function}
\label{sec:NLOcorr}
\medskip

In the last section we have analyzed the moments of the kinetic equations considered in the leading order of $\hbar$, which have led us to the hydrodynamic equations. Since the spin tensor enters with an extra power of $\hbar$, it is important to reexamine the Wigner function up to the next-to-leading order (NLO) in $\hbar$, where the functions ${\cal F}^{(1)}$  and ${\cal A}_{(1)}^\mu$ may be treated again as independent variables, and other coefficients are expressed in terms of these two functions and other NLO terms~\cite{Florkowski:2018ahw}:

\bel{eq:rP1a}
{\cal P}^{(1)}  = -\frac{1}{2m} \, \p^\mu  {\cal A}^{(0)}_\mu,
\eel
\bel{eq:rV1a}
{\cal V}^{(1)}_\mu &=& \frac{1}{m} \left(k_\mu {\cal F}^{(1)} 
- \frac{1}{2} \p^\nu {\cal S}^{(0)}_{\nu \mu} \right),
\eel
\bel{eq:rS1a}
{\cal S}_{\mu \nu}^{(1)} = \frac{1}{2m} \left(\p_\mu {\cal V}^{(0)}_\nu - \p_\nu {\cal V}^{(0)}_\mu \right) 
- \frac{1}{m} \epsilon_{\mu \nu \alpha \beta} k^\alpha {\cal A}_{(1)}^\beta.
\eel
Similarly to the leading order, the functions ${\cal F}_{(1)}$ and ${\cal A}^\nu_{(1)}$ satisfy the kinetic equations
\bel{eq:kineqF1}
k^\mu \p_\mu {\cal F}_{(1)}(x,k) = 0,
\eel
\bel{eq:kineqA1}
k^\mu \p_\mu {\cal A}^\nu_{(1)} (x,k) = 0, 
\quad k_\nu {\cal A}^\nu_{(1)} (x,k) = 0.
\eel
If ${\cal F}_{(1)}$ and ${\cal A}^\nu_{(1)}$ are determined, 
the quantities ${\cal P}^{(1)}$, ${\cal V}^{(1)}_\mu$, and ${\cal S}^{(1)}_{\mu \nu}$ are obtained from \EQSM{eq:rP1a}{eq:rS1a}.

An important aspect of the equations above is that the NLO coefficients gain non-trivial contributions from the leading order~\footnote{Note that \EQ{eq:eqforWC} with \EQn{eq:K} couples expressions that differ by one power of~$\hbar$.}. Hence, the equilibrium LO terms generate non-trivial corrections. Herein, we assume that ${\cal F}_{(1)} = {\cal A}^\nu_{(1)} = 0$, and the functions ${\cal P}^{(1)}$, ${\cal V}^{(1)}_\mu$, and ${\cal S}^{(1)}_{\mu \nu}$ are generated solely by the leading-order equilibrium terms:
\bel{eq:rP1aeq}
{\cal P}^{(1)}  = -\frac{1}{2m} \, \p^\mu  {\cal A}_{\rm eq, \mu},
\eel
\bel{eq:rV1aeq}
{\cal V}^{(1)}_\mu &=& -\frac{1}{2m} \p^\nu {\cal S}_{\rm eq, \nu \mu} ,
\eel
\bel{eq:rS1aeq}
{\cal S}_{\mu \nu}^{(1)} = \frac{1}{2m} \left(\p_\mu {\cal V}_{\rm eq,\nu} - \p_\nu {\cal V}_{\rm eq,\mu} \right) .
\eel
%

\subsection{Charge current}
\label{sec:chcur}
\medskip

Expressing the charge current in terms of the Wigner function $\Wxk$ we obtain~\cite{DeGroot:1980dk} 
\bea
N^\alpha_{\rm tot} (x) 
&=&  \trf \int d^4k \, \gamma^\alpha \, \Wxk
=   \int d^4k \, {\cal V}^\alpha (x,k).
\label{eq:Nalphacal1}
\eea
In the equilibrium case, to include the NLO corrections, besides \EQ{eq:VC1} we also use \EQ{eq:rV1aeq} to have a complete expression for ${\cal V}^\alpha(x,k)$. In this way we find
\bea
N^\alpha_{\rm tot} (x) &=&  N^\alpha_{\rm eq}(x) + \delta  N^\alpha_{\rm eq}(x) ,
\label{eq:Nalphacal2}
\eea
where
\bea
N^\alpha_{\rm eq} (x)  &=&   \frac{1}{m} \int d^4k \,  k^\alpha  {\cal F}_{\rm eq} (x,k) 
\label{eq:Nalpha}
\eea
and
\bea
\delta  N^\alpha_{\rm eq}(x) &=& - \frac{\hbar}{2m} \int d^4k \, \p_\lambda {\cal S}_{\rm eq}^{\lambda \alpha}(x,k) .
\label{eq:dNalpha}
\eea
One can easily check that \EQ{eq:Nalpha} agrees with \EQ{eq:Nalpha1}, and $ \p_\alpha \, \delta N^\alpha_{\rm eq}(x)  = 0$, which follows from the antisymmetry of the tensor ${\cal S}_{\rm eq}^{\lambda \alpha}(x,k)$. Hence, the conservation of the charge current \EQn{eq:Nalphacal1} agrees with the results obtained from the LO kinetic theory, discussed in \SEC{sect:pfspinkin}.

\subsection{GLW energy-momentum and spin  tensors} \label{sec:slmnGLW}
\medskip

Adopting the kinetic-theory framework derived by de Groot, van Leeuwen, and van Weert in \CIT{DeGroot:1980dk}, where the energy-momentum tensor is expressed directly by the trace of the Wigner function,  we can use the following expression
\bel{eq:tmunu1}
T^{\mu\nu}_{\rm GLW}(x)=\frac{1}{m}\trf \int d^4k \, k^{\mu }\,k^{\nu }\Wxk=\frac{1}{m} \int d^4k \, k^{\mu }\,k^{\nu } {\cal F}(x,k).
\eel
In the equilibrium case, we consider \EQ{eq:tmunu1} up to the first order in $\hbar$ using \EQ{eq:FC1} and setting ${\cal F}^{(1)}(x,k)=0$. In this way, we obtain the expressions that agree with \EQSM{Tmn}{prs}, hence $T^{\mu\nu}_{\rm GLW}(x) = T^{\mu\nu}_{\rm eq}(x)$. 

Let us turn now to the discussion of the spin tensor. In the GLW approach, it has the following form~\cite{DeGroot:1980dk}
\bel{eq:Smunulambda_de_Groot1}
\hspace{-0.5cm} S^{\lambda , \mu \nu }_{\rm GLW} =\frac{\hbar}{4} \!\int \!\! \di ^4k \, \trf \left[ \left( \left\{\sigma ^{\mu \nu },\gamma ^{\lambda }\right\}\!+\!\frac{2 i}{m}\left(\gamma ^{[\mu }k^{\nu ]}\gamma ^{\lambda }-\gamma ^{\lambda }\gamma ^{[\mu }k^{\nu ]}\right) \right) \Wxk \right]. 
\eel
For dimensional reasons, we have implemented here the Planck constant~$\hbar$. Its presence implies that in equilibrium we may take the leading order expression for the Wigner function and assume $ \Wxk=\Weqxk$. Using \EQ{eq:Weqpxk2} in \EQ{eq:Smunulambda_de_Groot1}, performing the appropriate traces over spinor indices, and then carrying out the integration over $k$ we obtain \EQ{eq:Smunulambda_de_Groot2}, hence $S^{\lambda , \mu \nu }_{\rm GLW}=S^{\lambda , \mu \nu }_{\rm eq}$.

The results obtained in this section are very much instructive, since they show us that the GLW formulation appears naturally in the context of the kinetic theory --- the conservation laws are obtained as the zeroth and first moments of the kinetic equations fulfilled by the scalar and axial-vector coefficients. Moreover, since the GLW energy-momentum tensor is symmetric, the GLW spin tensor should be conserved separately. Hence, the perfect-fluid hydrodynamics can be obtained from the equations
\bea
\p_\alpha N^\alpha_{\rm eq}(x)  = 0, \quad \p_\alpha T^{\alpha\beta}_{\rm GLW}(x) = 0, \quad
\p_\lambda S^{\lambda , \mu \nu }_{\rm GLW}(x) = 0 .
\label{eq:GLWhydro}
\eea
Equations (\ref{eq:GLWhydro}) can be interpreted as a specific realization of the general framework based on \EQS{TSj3}.

\subsection{Canonical tensors} 
\label{sec:slmnCAN}
\medskip

The canonical forms of the energy-momentum and spin tensors, $\TmnU_{\rm can}(x)$ and $S^{\lambda , \mu \nu }_{\rm can}(x)$, can be obtained directly from the Dirac Lagrangian by applying the Noether theorem~\cite{Itzykson:1980rh}:
\bel{eq:tmunu1can1}
T^{\mu\nu}_{\rm can}(x)= \int d^4k \,k^{\nu } {\cal V}^\mu(x,k)
\eel
and
\bea
\hspace{-0.75cm} S^{\lambda , \mu \nu }_{\rm can}(x)\!\!\!&=&\!\!\!\frac{\hbar}{4}\!  \int \!d^4k \,\trf \left[ \left\{\sigma ^{\mu \nu },\gamma ^{\lambda }\right\} \Wxk  \right] 
= \frac{\hbar}{2} \epsilon^{\kappa \lambda \mu \nu} \int \!d^4k \, {\cal A}_{ \kappa}(x,k).
\label{eq:Smunulambda_canonical1}
\eea
Here we have used the anticommutation relation $\left\{\sigma ^{\mu \nu },\gamma ^{\lambda }\right\} = -2 \epsilon^{\mu\nu\lambda\kappa} \gamma_\kappa \gamma_5$ to express directly the canonical spin tensor by the axial-vector coefficient function ${\cal A}_{\kappa}(x,k)$. 

Including the components of ${\cal V}^\mu(x,k)$ up to the first order in the equilibrium case we obtain
\bel{eq:tmunu1can2}
T^{\mu\nu}_{\rm can}(x) = T^{\mu\nu}_{\rm GLW}(x) + \delta T^{\mu\nu}_{\rm can}(x) 
\eel
where
\bel{deltaTmunu}
\delta T^{\mu\nu}_{\rm can}(x)  = -\frac{\hbar}{2m} \int d^4k k^\nu \partial_\lambda {\cal S}^{\lambda \mu}_{\rm eq}(x,k) = -\partial_\lambda S^{\nu , \lambda \mu }_{\rm GLW}(x).
\eel
The canonical energy-momentum tensor should be exactly conserved, hence, in analogy to \EQ{eq:Tcon} we require 
\bel{eq:Tconcan}
\p_\alpha T^{\alpha\beta}_{\rm can}(x) = 0.
\eel
It is interesting to observe that \EQSTWO{eq:Tcon}{eq:Tconcan} are consistent, since $\partial_\mu \, \delta T^{\mu\nu}_{\rm can}(x) = 0$. The latter property follows directly from the definition of $\delta T^{\mu\nu}_{\rm can}(x) $, see \EQ{deltaTmunu}.

For the equilibrium spin tensor it is enough to consider the axial-vector component in \EQ{eq:Smunulambda_canonical1}  in the zeroth order, $ {\cal A}^{(0)}_{ \kappa}(x,k)= {\cal A}_{\rm eq,  \kappa}(x,k)$. Then, using \EQ{eq:AEeqpm} in \EQ{eq:Smunulambda_canonical1} and carrying out the integration over the four-momentum $k$ we get
\bea
\hspace{-0.75cm}S^{\lambda , \mu \nu }_{\rm can}  &=& \frac{\hbar \sinh(\zeta)\cosh(\xi)}{\zeta} \int dP \, e^{-\beta \cdot p}\left(\omega ^{\mu \nu } p^{\lambda}+\omega ^{\nu \lambda } p^{\mu}+\omega ^{\lambda \mu } p^{\nu}\right) \nn \\
\hspace{-0.75cm}&=& \frac{\hbar w}{4 \zeta} \left( u^\lambda \omega^{\mu\nu} + 
u^\mu \omega^{\nu \lambda} + u^\nu \omega^{\lambda \mu}
\right) = S^{\lambda , \mu \nu }_{\rm GLW} + S^{\mu , \nu \lambda }_{\rm GLW}+ S^{\nu , \lambda \mu }_{\rm GLW}.
\label{eq:Smunulambda_canonical2}
\eea
It is interesting to notice that the energy-momentum tensor \EQn{eq:tmunu1can2} is not symmetric. In such a case, the spin tensor is not conserved and its divergence is equal to the difference of the energy-momentum components. For the case discussed in this section we obtain
\bea
\p_\lambda S^{\lambda , \mu \nu }_{\rm can}(x) = T^{\nu\mu}_{\rm can} - T^{\mu\nu}_{\rm can}
= -\partial_\lambda S^{\mu , \lambda \nu }_{\rm GLW}(x) + \partial_\lambda S^{\nu , \lambda \mu }_{\rm GLW}(x). 
\label{eq:Scancon}
\eea

One can immediately check, using the last line of \EQ{eq:Smunulambda_canonical2}, that \EQ{eq:Scancon} is consistent with the conservation of the spin tensor in the GLW approach.  In fact, it leads to exactly the same set of differential equations as the conservation of the GLW spin tensor. Since the conservation of the energy and linear momentum for the GLW and canonical versions leads also to the same differential equations, we conclude that the GLW and canonical formulations determine the same dynamics of hydrodynamic variables. The differences between two formulations reside, however, in the allocation of densities (of various physical quantities such as energy, linear momentum, and angular momentum). The canonical expressions for densities depend on gradients of hydrodynamic variables, which make them more difficult to interpret. On the other hand, the GLW densities depend on the hydrodynamic variables alone. Hence it is easy to start with the GLW formulation and obtain the corresponding canonical results afterwards.

\subsection{Connecting GLW and canonical formulations} \label{sec:PsG}
\medskip

In the last section we have discussed the energy-momentum and spin tensors obtained from the canonical formalism and related them to the expressions introduced by de Groot, van~Leeuven, and van~Weert.  In this section we demonstrate that the two versions of the energy-momentum and spin tensors are connected by a pseudo-gauge transformation. Indeed, if we introduce the tensor $\Phi_{\rm can}^{\lambda, \mu\nu}$ defined by the relation
\bel{Phi}
\Phi_{\rm can}^{\lambda, \mu\nu} 
\equiv S^{\mu , \lambda \nu }_{\rm GLW}
-S^{\nu , \lambda \mu }_{\rm GLW},
\eel
we can write
\bel{psg1GLW}
S^{\lambda , \mu \nu }_{\rm can}= S^{\lambda , \mu \nu }_{\rm GLW} -\Phi_{\rm can}^{\lambda, \mu\nu}
\eel
and
\bel{psg2GLW}
T^{\mu\nu}_{\rm can} = T^{\mu\nu}_{\rm GLW} + \frac{1}{2} \p_\lambda \left(
\Phi_{\rm can}^{\lambda, \mu\nu}
+\Phi_{\rm can}^{\mu, \nu \lambda} 
+ \Phi_{\rm can}^{\nu, \mu \lambda} \right).
\eel
Here, we have used the property that both $S^{\lambda , \mu \nu }_{\rm GLW} $ and $\Phi_{\rm can}^{\lambda, \mu\nu} $ are antisymmetric with respect to exchange of the last two indices. Equations \EQn{psg1} and \EQn{psg2} are an example of the pseudo-gauge transformation introduced and discussed in Sec.~\ref{sect:pseudogauget}. The explicit form of the pseudogauge transformation that connects the GLW and canonical frameworks, defined in terms of the Dirac fields $\Psi$, is given in Ch. IV, Sec. 1 of Ref.~\cite{DeGroot:1980dk}.

The most common use of such a transformation is connected with a change from the canonical formalism to the Belinfante one \cite{Belinfante:1940} --- it provides a symmetric energy-momentum tensor and eliminates completely the spin tensor.  Using~\EQ{belinf} we explicitly find $T^{\mu\nu}_{\rm Bel} = T^{\mu\nu}_{\rm GLW}- \frac{1}{2} \partial_\lambda \left(S^{\nu,\lambda\mu}_{\rm GLW}+ S^{\mu,\lambda\nu}_{\rm GLW}\right)$. This formula implies that $\partial_\mu T^{\mu\nu}_{\rm Bel} = \partial_\mu T^{\mu\nu}_{\rm GLW}$, due to the conservation of the GLW spin tensor. Consequently, the Belinfante and GLW energy-momentum tensors lead to the same equations for hydrodynamic variables, but to get the same results a special tuning of the initial conditions should be made (see the similar comment at the end of the previous section, where we compare the canonical and GLW expressions). 

Interestingly, the leading-order, classical expressions for the canonical, GLW, and the Belinfante energy-momentum tensors are all the same. The main difference between the Belinfante and the other two approaches is that no spin tensor is available in the Belinfante case, hence, the conservation of the Belinfante energy-momentum tensor is not sufficient to determine the space-time evolution of the polarization tensor $\omega_{\mu\nu}$. It should be determined by other physical conditions. One option is to directly relate $\omega_{\mu\nu}$ to the thermal vorticity. Within the processes discussed in this work we cannot select a mechanism responsible for such an identification. Most likely, these two tensors become equal as a result of dissipative phenomena that remain beyond the perfect-fluid picture analyzed herein.

It has been recently argued that the use of tensors that differ by the pseudo-gauge transformation leads to different predictions for measurable quantities such as spectrum and polarization of particles~\cite{Becattini:2018duy}. We expect that the results presented in this work can be useful to study  such effects in more detail within explicitly defined hydrodynamic models. In particular, our discussion above sheds new light on the role of the initial conditions --- they should be properly adjusted when tensors differing by a pseudogauge transformation are compared.

\subsection{FFJS hydrodynamic model} \label{sec:FFJS}
\medskip

Connections between different hydrodynamic models with spin, realized by the pseudo-gauge transformations, can be used to clarify the role of the FFJS hydrodynamic model~\CITn{Florkowski:2017ruc} which introduced for the first time the concept of the spin chemical potential as the Lagrange multiplier that can play a role of a hydrodynamic variable. The FFJS approach employs the energy-momentum tensor in the GLW version and the phenomenological spin tensor of the form
\bea 
S^{\lambda , \mu \nu }_{\rm ph} = \frac{\hbar w}{4 \zeta}
u^\lambda \omega^{\mu\nu}.
\label{eq:Sph1}
\eea
It can be related to the canonical and GLW spin tensors through the following relations
\bea 
S^{\lambda , \mu \nu }_{\rm can} = S^{\lambda , \mu \nu }_{\rm ph}
+ S^{\mu , \nu \lambda }_{\rm ph} + S^{\nu , \lambda \mu }_{\rm ph}
\label{eq:Sph2}
\eea
\bea 
S^{\lambda , \mu \nu }_{\rm GLW} = S^{\lambda , \mu \nu }_{\rm ph}+ S^{\lambda , \mu \nu }_\Delta,
\label{eq:Sph3}
\eea
where the tensor $S^{\lambda , \mu \nu }_\Delta$ is defined by \EQ{eq:Smunulambda_de_Groot2}.

Equation \EQn{eq:Sph3} can be interpreted as an element of the pseudo-gauge transformation leading from the canonical formalism to the phenomenological FFJS model~\CITn{Florkowski:2017ruc}, 
\bea 
S^{\lambda , \mu \nu }_{\rm ph} = S^{\lambda , \mu \nu }_{\rm can}
- \Phi_{\rm ph}^{\lambda , \mu \nu }
\label{eq:Sph4}
\eea
with
\bea 
\Phi_{\rm ph}^{\lambda , \mu \nu }
= S^{\mu , \nu \lambda }_{\rm ph} + S^{\nu , \lambda \mu }_{\rm ph}.
\label{eq:Phi-ph}
\eea
It is interesting to check now the form of the energy-momentum tensor, which is induced by the superpotential defined above (in the transition from the canonical forms). According to \EQ{psg1} we use
\bel{eq:Tmnph1}
T^{\mu\nu}_{\rm ph} = T^{\mu\nu}_{\rm can} + \frac{1}{2} \p_\lambda \left(
\Phi_{\rm ph}^{\lambda, \mu\nu}
+\Phi_{\rm ph}^{\mu, \nu \lambda} 
+ \Phi_{\rm ph}^{\nu, \mu \lambda} \right)
\eel
and obtain
\bel{eq:Tmnph2}
T^{\mu\nu}_{\rm ph} = T^{\mu\nu}_{\rm GLW} -
\p_\lambda S^{\nu , \lambda \mu }_{\Delta}.
\eel
Since $S^{\nu , \lambda \mu }_{\Delta}=-S^{\nu , \mu \lambda}_{\Delta}$, the conservation of the phenomenological tensor $T^{\mu\nu}_{\rm ph}$ coincides with the conservation of $T^{\mu\nu}_{\rm GLW}$, which validates the use of the equation $\p_\mu T^{\mu\nu}_{\rm GLW} = 0$ in \CIT{Florkowski:2017ruc}. 

Since $T^{\mu\nu}_{\rm ph}$ defined by~\EQ{eq:Tmnph2} is not symmetric, the spin tensor~\EQn{eq:Sph1} is in general not conserved. Similarly to \EQn{Sdiv} we should have
\bea
\p_\lambda S^{\lambda , \mu \nu }_{\rm ph} = 
T^{\nu\mu}_{\rm ph}-T^{\mu\nu}_{\rm ph},
\label{eq:Sphcon1}
\eea 
which after simple manipulations yields
\bea
\p_\lambda S^{\lambda , \mu \nu }_{\rm ph} = 
- \p_\lambda S^{\lambda , \mu \nu }_{\Delta}.
\label{eq:Sphcon2}
\eea 
One can easily notice that the equation above defines the conservation of the GLW spin tensor. In the hydrodynamic model~\CITn{Florkowski:2017ruc} it is assumed that the spin tensor $S^{\lambda , \mu \nu }_{\rm ph} $ is conserved, which is equivalent to neglecting the term on the right-hand side of \EQ{eq:Sphcon2}.~\footnote{The term $S^{\lambda , \mu \nu }_{\Delta}$ may be interpreted as a relativistic correction to the term $S^{\lambda , \mu \nu }_{\rm ph}$. In the case of heavy particles, such as the $\Lambda$ hyperon, this could be a reasonable approximation (to be verified by future numerical calculations). } The improvement of the FFJS model defined in this section is straightforward and leads to a framework that is consistent with both the GLW and canonical formulations. 

At this point it is important to emphasize that the GLW and canonical formulations of relativistic hydrodynamics with spin (as well as the phenomenological formulation modified along the lines described herein) yield the same values of the total energy, total linear momentum, and total angular momentum. This is so, since they are all connected by pseudogauge transformations explicitly defined above.

\subsection{Pauli-Luba\'nski vector} \label{sec:PL0}
\medskip

Starting from the definition of the  Pauli-Luba\'nski (PL) four-vector in the form $\Pi_\mu=-\f{1}{2} \epsLmnab J^{\nu\alpha}p^\beta$ 
(where $J^{\nu\alpha}$ is the total angular momentum), and following Ref.~\cite{Becattini:2013fla}, we introduce the phase-space density of $\Pi_\mu$ defined by the following expression
\bel{PL10}
E_p \f{d \Delta \Pi_\mu(x,p)}{d^3p}  = -\f{1}{2} \epsLmnab \, \Delta \Sigma_\lambda(x) \, 
E_p \f{d J^{\lambda, \nu\alpha}(x,p)}{d^3p}
\f{p^\beta}{m}.
\eel
Here $\Delta \Sigma_\lambda(x)$ denotes an element of the hyper-surface containing the particles of interest, and $E_p d J^{\lambda, \nu\alpha}(x,p)/d^3p$ denotes the invariant angular momentum phase-space density of particles with four-momentum $p$. One can notice that the density $E_p d J^{\lambda, \nu\alpha}/d^3p$ may be replaced by $E_p d S^{\lambda, \nu\alpha}/d^3p$, since these two differ by terms proportional to four-momenta that do not contribute to \EQ{PL10},
\bel{PL11}
E_p \f{d \Delta \Pi_\mu(x,p)}{d^3p}  = -\f{1}{2} \epsLmnab \, \Delta \Sigma_\lambda(x) \, 
E_p \f{d S^{\lambda, \nu\alpha}(x,p)}{d^3p}
\f{p^\beta}{m}.
\eel
Using the GLW definition of the spin tensor we find
\bea
\f{1}{2} \epsLmnab
E_p \f{d S^{\lambda, \nu\alpha}(x,p)}{d^3p} &=&  \, \f{\hbar \cosh(\xi)}{(2\pi)^3} \, 
\f{ \sinh(\zeta)}{ \zeta} \, e^{- p \cdot \beta} p^\lambda {\tilde \omega}_{\mu \beta}.
\label{Slna2}
\eea
It should be emphasized at this point, however, that exactly the same result is obtained for the right-hand-side of Eq.~(\ref{Slna2}) if one uses the canonical definition of the spin tensor or the phenomenological one. This property is explicitly demonstrated in Ref.~\cite{Florkowski:2017dyn}.

Since we are interested in the polarization effect per particle, it is necessary to introduce the particle density in the volume $\Delta \Sigma$. It is described by the expression 
\bea
E_p \f{d \Delta {\cal N}}{d^3p} &=& 
= \f{4}{(2\pi)^3}   \, \Delta \Sigma \cdot  p \, e^{-p \cdot \beta} \cosh(\xi) \cosh(\zeta).
\label{DcalN}
\eea
The PL vector per particle is then obtained by dividing \EQ{PL10} by \EQ{DcalN},
\bel{PL4}
\pi_\mu(x,p) \equiv \f{\Delta \Pi_\mu(x,p)}{\Delta {\cal N}(x,p)} &=& 
-\f{\hbar \tanh(\zeta)}{4 m \zeta} \, {\tilde \omega}_{\mu \beta} \, p^\beta.
\eel

In order to transform the four-vector $\pi^\mu$ to the local rest frame of a particle with momentum $p$, we use the canonical boost. In this way, we can express the time and space components of $\pi^\mu = (\pi^0, \piv)$ in the LAB frame in the three-vector notation:
\bel{pi0PRF}
\pi^0_\ast   = 0
\eel
and
\bel{pivPRF}
\piv_\ast   =   - \f{\hbar \tanh(\zeta)}{4 \zeta}  \Pv \, .
\eel
This is a crucial result for the framework presented in this work, showing that the space part of the PL vector in PRF agrees with the mean polarization three-vector obtained in Sec.~\ref{sec:P3V} (note that $\hbar$ should be installed therein). Equation (\ref{pivPRF}) legitimizes the use of the GLW and canonical forms for the spin tensor in the expression for the Pauli-Luba\'nski vector, as fully consistent with a direct calculation of the spin polarization from the spin distribution functions via Eq.~(\ref{eq:avPv1}).

\section{Classical treatment of spin  \label{chapt:classspin}}


\subsection{Internal angular momentum tensor}
\label{sec:int-ang-mom-ten}
\medskip

In the classical treatment of spin one introduces the internal angular momentum tensor $s^{\alpha\beta}$ \CITn{Mathisson:1937zz} defined here in terms of the particle four-momentum $p_\gamma$ and spin four-vector~$s_\delta$~\cite{Itzykson:1980rh},
\bea
s^{\alpha\beta} = \f{1}{m} \epsUabgd p_\gamma s_\delta.
\label{eq:sab}
\eea
Equation \EQn{eq:sab} implies that $s^{\alpha\beta} = -s^{\beta\alpha}$ and
\bea
p_\alpha s^{\alpha\beta}  = 0.
\label{eq:frenkel}
\eea
The last condition is known in the literature as the Frenkel (or Weyssenhoff) condition. The spin four-vector is orthogonal to four-momentum
\bea
s \cdot p = 0,
\label{eq:sp}
\eea
hence, we can invert \EQ{eq:sab} to obtain
\bea
s^{\alpha} = \f{1}{2m} \epsUabgd p_\beta s_{\gamma \delta}.
\label{eq:sabinv}
\eea
In PRF, where $p^\mu = (m,0,0,0)$, the four-vector $s^\alpha$ has only space components, $s^\alpha = (0,\sv_*)$, with the normalization $\spinl = \spin$. For spin one-half particles we use the value of the Casimir operator,
\bea
\spin^2 = \f{1}{2} \left( 1+ \f{1}{2}  \right) = \f{3}{4}.
\label{eq:spin2}
\eea

\subsection{Invariant measure in spin space}
\label{sec:inv-mes-spin}
\medskip

A straightforward generalization of the phase-space distribution function $f(x,\pv)$ is a spin dependent distribution $f(x,\pv,s)$\footnote{A classical spin dependent distribution function was considered also in Ref.~\cite{Chen:2013iga}.}. Different orientations of spin can be integrated out with the help of a covariant measure
\bea
\int dS \ldots = \f{m}{\pi \spin}  \, \int d^4s \, \delta(s \cdot s + \spin^2) \, \delta(p \cdot s) \ldots \,.
\label{eq:dS}
\eea
The two delta functions in \EQ{eq:dS} take care of the normalization and orthogonality conditions, while the prefactor $m/(\pi \spin)$  is chosen to obtain the normalization
\bea
\int dS = \f{m}{\pi \spin}  \int \, d^4s \, \delta(s \cdot s + \spin^2) \, \delta(p \cdot s) = 2,
\label{eq:dSint}
\eea
that accounts for the spin degeneracy factor of spin one-half particles.

\subsection{Equilibrium distribution}
\label{sec:eq-f-spin}
\medskip

To construct the equilibrium function for particles with spin we have to identify the so-called collisional invariants of the Boltzmann equation. In addition to four-momentum and conserved charges, for particles with spin one can include the total angular momentum
\bea
j_{\alpha\beta} = l_{\alpha\beta} +s_{\alpha\beta} = x_\alpha p_\beta - x_\beta p_\alpha +s_{\alpha\beta}.
\label{eq:jls}
\eea
If we change the reference point for the calculation of the orbital part by replacing $x^\mu$ by $x^\mu + \delta^\mu$ we obtain the new orbital angular momentum $l^\prime_{\alpha\beta} = l_{\alpha\beta} + \delta_\alpha p_\beta - \delta_\beta p_\alpha$. At face value, this change could be absorbed into a new definition of the spin part, $s^\prime_{\alpha\beta} = s_{\alpha\beta} - \delta_\alpha p_\beta + \delta_\beta p_\alpha$, leaving the total angular momentum $j_{\alpha\beta}$ unchanged. However, the Frenkel condition \EQn{eq:frenkel} applied to the old and new spin parts implies that $\delta_\beta =  p_\beta (p \cdot \delta)/m^2$ in this case, hence $s^\prime_{\alpha\beta} = s_{\alpha\beta}$. Consequently, for massive particles the changes of the orbital part of the angular momentum cannot be compensated by a redefinition of the spin part if \EQSTWO{eq:sab}{eq:frenkel} are used. 

The locality of the standard Boltzmann equation~\footnote{By this we mean here the dependence of the collsion term on a single space-time component $x$ that may be set equal to zero by a translation. A non-local version of the Boltzmann equation was proposed, for example, in~\CIT{Jaiswal:2012qm} where non-local effects are included through gradients of the phase-space distribution function.} suggests that the orbital part in \EQ{eq:jls} can be eliminated, and the spin part can be considered separately. For elastic binary collisions of particles $1$ and $2$ going to $1^\prime$ and $2^\prime$, this suggests that~\CITn{Weert:1970aa}
\bea
s^{\alpha\beta}_1 + s^{\alpha\beta}_2 = s^{\alpha\beta}_{1^\prime} + s^{\alpha\beta}_{2^\prime} \, .
\label{eq:sab-con}
\eea

The tensors $s^{\alpha \beta}$ appearing in \EQ{eq:sab-con} depend on the four-momenta $p$ and  spin four-vectors $s$ of colliding particles. To examine more closely the properties which a collision process should have in order to satisfy \EQ{eq:sab-con}, we switch to the center-of-mass frame of the colliding particles 1 and 2 (CMS). In this frame the incoming three-momenta are $\pv_1 = \pv$ and  $\pv_2 = -\pv$. After the collision, the three-momenta of outgoing particles are $\pv_{1^\prime} = \pv^\prime$ and  $\pv_{2^\prime} = -\pv^\prime$. Since we deal with an elastic collision,  $|\pv|=|\pv^\prime|$ and $E_p = E_{p^\prime} = E$. The three-velocities before and after the collision are denoted as $\vv = \pv/E$ and $\vv^\prime = \pv^\prime/E$. Using this notation, the six independent equations included in the tensor equation \EQn{eq:sab-con} may be written as two vector equations:
\bea
E \LB \sv_1 + \sv_2 \RB - \pv \LB s^0_1 - s^0_2 \RB &=& 
E \LB \sv_{1^\prime} + \sv_{2^\prime} \RB - \pv^\prime  \LB s^0_{1^\prime} - s^0_{2^\prime} \RB,  \label{eq:el1} \\
\pv \times \LB \sv_1 - \sv_2 \RB &=&
\pv^\prime \times \LB
\sv_{1^\prime} - \sv_{2^\prime} \RB. 
\label{eq:el2}
\eea 
Here we have introduced the time and space components of the spin four-vectors, $s^\mu_1 = (s^0_1, \sv_1)$, $s^\mu_2 = (s^0_2, \sv_2)$, $s^\mu_{1^\prime} = (s^0_{1^\prime}, \sv_{1^\prime})$, and $s^\mu_{2^\prime} = (s^0_{2^\prime}, \sv_{2^\prime})$. The orthogonality condition \EQn{eq:sp} implies that: $s_1^0 = \vv \cdot \sv_1$, $s_2^0 = -\vv \cdot \sv_2$, $s_{1^\prime}^0 = \vv^\prime \cdot \sv_{1^\prime}$, and $s_{2^\prime}^0 = -\vv^\prime \cdot \sv_{2^\prime}$. Hence, \EQSTWO{eq:el1}{eq:el2} are equivalent to
\bea
\sv_1 + \sv_2  - \vv 
\LSB \vv \cdot \LB \sv_1 + \sv_2 \RB \RSB &=& 
\sv_{1^\prime} + \sv_{2^\prime}  - \vv^\prime  
\LSB \vv^\prime \cdot \LB \sv_{1^\prime} + \sv_{2^\prime} \RB \RSB, \label{eq:el1a}  \\
\vv \times \LB \sv_1 - \sv_2 \RB &=&
\vv^\prime \times \LB
\sv_{1^\prime} - \sv_{2^\prime} \RB. 
\label{eq:el2a}
\eea 
It is interesting to notice that \EQSTWO{eq:el1a}{eq:el2a} admit two types of simple solutions;  they correspond to the cases where either the sum of two spin three-vectors or their difference (before and after the collision) vanishes. With a grain of salt, they may be interpreted as collisions in the spin singlet and triplet states. 
\begin{figure}
\centering
\includegraphics[width=0.45\textwidth]{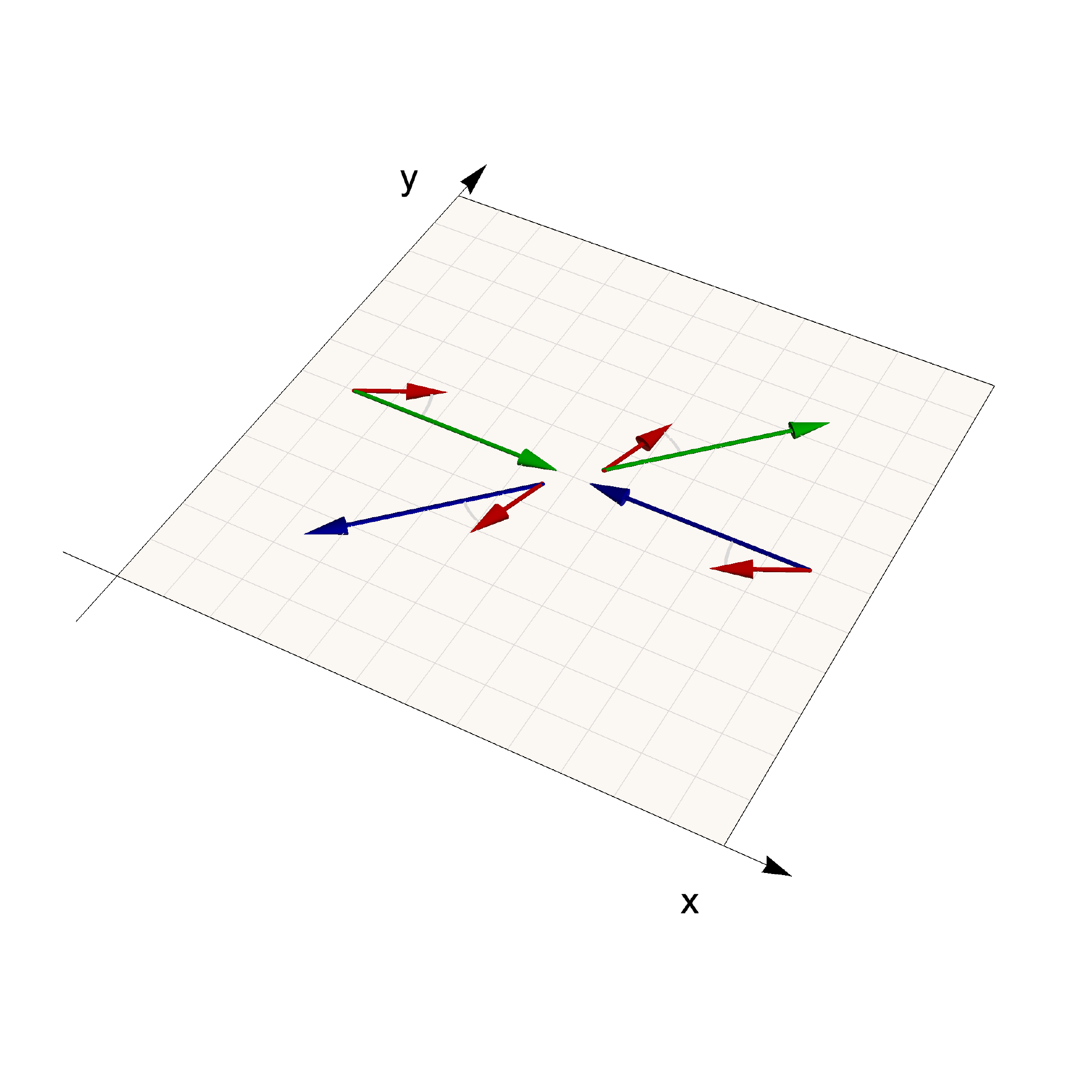}
\caption{(Color online) Schematic view of the collision of classical spins which have non-zero components along the momenta of colliding particles in the center-of-mass frame, see \EQSTWO{eq:el2bis}{eq:norm_s_1}. }
\label{fig:s1}
\end{figure}
\begin{figure} 
\centering
\includegraphics[width=0.45\textwidth]{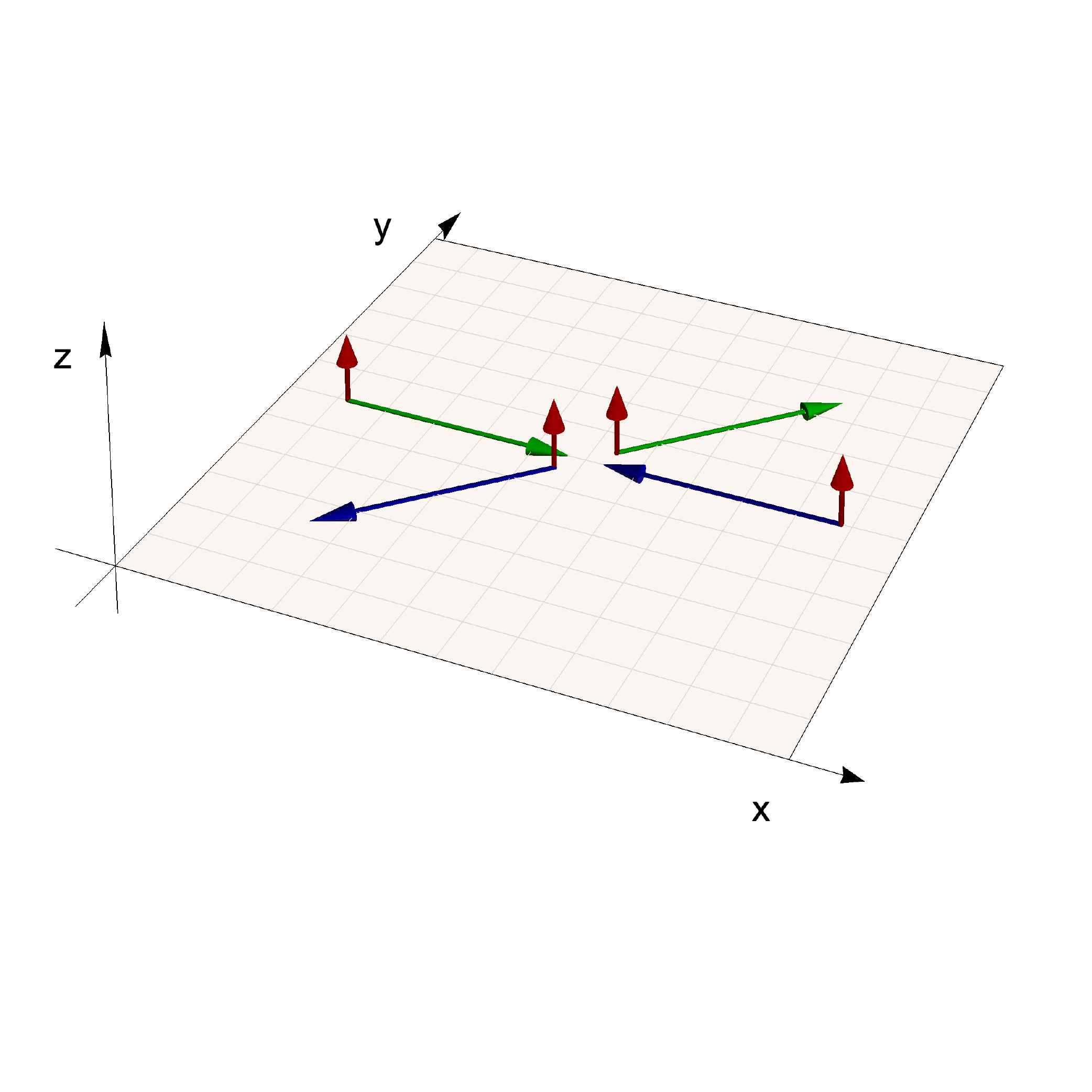}
\caption{(Color online) Schematic view of the collision of classical spins which are orthogonal to the collision plane in the center-of-mass frame, see \EQSTWO{eq:elabis}{eq:norm_s_2}.}
\label{fig:s2}
\end{figure}
\begin{itemize} 
	\item[{\bf 1.}] If the sum of initial and final spin three-vectors is zero, we may write that $\sv_1 = - \sv_2 = \sv$ and $\sv_{1^\prime} = - \sv_{2^\prime} =  \sv^\prime$. In this case \EQ{eq:el1a} is automatically fulfilled, while \EQ{eq:el2a} becomes
	\bea 
	\vv \times \sv = \vv^\prime \times \sv^\prime.
	\label{eq:el2bis}
	\eea 
	The structure of this equation implies that all vectors appearing in \EQn{eq:el2bis} lie in the same plane, and a natural solution for $\sv^\prime$ is a vector obtained from $\sv$ by exactly the same rotation that transforms $\vv$ into $\vv^\prime$. In this case the time component of the spin vector is the same for all four particles participating in the collision, $s_0 = \vv \cdot \sv$, and the normalization of $\sv$ is obtained from the equation
	\bea 
	| \sv | = \f{\spin}{\sqrt{1 - v^2 \cos^2\phi}},
	\label{eq:norm_s_1}
	\eea 
	where $\phi$ is the angle between the three-vectors $\vv$ and $\sv$. Note that this case includes a situation in which the spin vectors are parallel to velocities ($\phi=0$ and $| \sv | = \spin/\sqrt{1-v^2}$).
	
	\item[{\bf 2.}]
	If the difference of initial and final spin three-vectors is zero, we may write that $\sv_1 =  \sv_2 = \sv$ and $\sv_{1^\prime} = \sv_{2^\prime} =  \sv^\prime$. In this case \EQ{eq:el2a} is automatically fulfilled, whereas \EQ{eq:el1a} takes the form
	\bea 
	\sv   - \vv 
	\LB \vv \cdot \sv \RB &=& 
	\sv^\prime   - \vv^\prime  
	\LB \vv^\prime \cdot  \sv^\prime \RB.
	\label{eq:elabis}
	\eea 
	To solve this equation we assume that
	$\sv^\prime = \sv$, and the vector $\sv$ is perpendicular to both $\vv$ and $\vv^\prime$. In this case $s_0 = 0$ and
	\bea 
	| \sv | = \spin.
	\label{eq:norm_s_2}
	\eea 
\end{itemize}

If the collision integral allows for processes such as discussed above, the conservation law \EQn{eq:sab-con} should be definitely included among the other, more common, conservation laws. This implies that we can introduce a spin-dependent, equilibrium distribution functions for particles and antiparticles in the form
\bea
f^\pm_{\rm eq}(x,p,s) = \exp\LB - p \cdot \beta(x) \pm \xi(x) + \f{1}{2}  \omega_{\alpha \beta}(x) s^{\alpha\beta} \RB.
\label{eq:fpm-spin}
\eea
As in the previous sections, the tensor $\omega_{\alpha \beta}(x)$ plays a role of the chemical potential conjugated to the spin angular momentum and we keep calling it the spin polarization tensor. We note that a similar procedure was applied before in \CIT{Czyz:1986mr} to introduce colored chemical potentials into the kinetic theory of quark-antiquark plasma with classical description of the color charge. We also note at this point that the Frenkel condition \EQn{eq:frenkel} imposed on the tensor $s^{\alpha\beta}$ has no effect on the spin polarization tensor $\omega_{\alpha\beta}$. In particular we do not demand that $u_\mu \omega^{\mu\nu}  =0$. This constraint is sometimes also called the Frenkel condition but, in general, it is not fulfilled (see, for example, the case of global thermodynamic equilibrium studied in \CIT{Becattini:2009wh}).

Using \EQ{eq:sab} we write 
\bea
\f{1}{2}  \omega_{\alpha \beta} s^{\alpha\beta} = \f{p_\gamma}{m} \, {\tilde \omega}^{\gamma \delta}
s_\delta.
\label{eq:bspin}
\eea
In PRF, the components of the four-vector $(p_\gamma/m)  {\tilde \omega}^{\gamma \delta}$ reproduce the elements of the polarization three-vector, see Eqs.~(\ref{AvebPRF}), (\ref{PL4}) and (\ref{pivPRF}), hence, we can write
\bea
\f{1}{2}  \omega_{\alpha \beta} s^{\alpha\beta}  =  \bv_* \cdot \sv_* =  \Pv \cdot \sv_* \,.
\eea
This formula will be frequently used below.

\subsection{Charge current and energy-momentum tensor}
\label{sec:charge-spin}
\medskip

The charge current is obtained from the straightforward generalization of the standard definition
\bea
N^\mu_{\rm eq} = \int dP   \int dS \, \, p^\mu \, \left[f^+_{\rm eq}(x,p,s)-f^-_{\rm eq}(x,p,s) \right],
\label{eq:Neq-sp0}
\eea
which after using the forms of the equilibrium functions \EQn{eq:fpm-spin} leads to the expression
\bea
N^\mu_{\rm eq} = 2 \sinh(\xi) \int dP \, p^\mu \, \exp\LB- p \cdot \beta \RB
\int dS \, \exp\LB \f{1}{2}  \omega_{\alpha \beta} s^{\alpha\beta} \RB.
\label{eq:Neq-sp1}
\eea
For small values of the polarization tensor $\omega$, one can expand the exponential function, which gives
\bea
N^\mu_{\rm eq} &=& 2 \sinh(\xi) \int dP \, p^\mu \, e^{- p \cdot \beta} \int dS \, \left[1 +  \f{1}{2}  \omega_{\alpha \beta} s^{\alpha\beta}\right] \nn \\
&=& 4 \sinh(\xi) \int dP \, p^\mu \, e^{- p \cdot \beta} . 
\label{eq:Neq-sp2}
\eea
Here we used the normalization \EQn{eq:dSint} and the fact that $ \int dS \, s^\mu = 0$. Equation \EQn{eq:Neq-sp2} agrees up to the first order in $\omega$ with \EQ{Nmu}.

The treatment of the energy-momentum tensor is analogous. Repeating the same steps we find
\bea
T^{\mu \nu}_{\rm eq} &=& \int dP   \int dS \, \, p^\mu p^\nu \, \left[f^+_{\rm eq}(x,p,s) + f^-_{\rm eq}(x,p,s) \right] \nn \\
&=& 2 \cosh(\xi) \int dP \, p^\mu p^\nu \, \exp\LB - p \cdot \beta\RB
\int dS \, \exp\LB  \f{1}{2}  \omega_{\alpha \beta} s^{\alpha\beta} \RB,
\label{eq:Neq-sp01}
\eea
which agrees with \EQ{Tmn} again up to the first order in $\omega$.

In order to obtain the last integral in \EQ{eq:Neq-sp01} for arbitrary value of $\omega$ we switch to PRF,
\bea
\int dS \, \exp\LB  \f{1}{2}  \omega_{\alpha \beta} s^{\alpha\beta}\RB 
&=& \f{m}{\pi \spin} \int d^4s \, \delta(s \cdot s + \spin^2) \, \delta(p \cdot s)
\exp\LB \,  \f{1}{2}  \omega_{\alpha \beta} s^{\alpha\beta} \RB \nn \\
&& \hspace{-3cm} = \f{m}{\pi \spin} \int ds_0 \int \spinl^2 d\spinl \int d\Omega \, \delta(\spinl^2 - \spin^2) \, \delta(m s_0) e^{ \Pv \cdot  \sv_*} .
\label{eq:spinint1}
\eea
Here $d\Omega=\sin(\theta) d\theta d\phi$ denotes the integration over the solid angle (two independent directions of the three-vector $\sv_*$). Doing the integrals over $s_0$, $\spinl$ and $\phi$, and using the substitution $x =\cos(\theta)$, we obtain 
\bea
\int dS \, \exp\LB  \f{1}{2}  \omega_{\alpha \beta} s^{\alpha\beta}\RB  
&=& \int\limits_{-1}^{+1} e^{ \spin \,P  \,x} dx = \f{2 \sinh( \spin P)}{ \spin \, P},
\label{eq:spinint2}
\eea
where $P = |\Pv| =  |\bv_*|$. As expected, for small values of $P$ we reproduce the normalization result~\EQn{eq:dSint}. Interestingly, since $P$ depends on momentum, the energy-momentum tensor for large values of $\omega$ has no longer a simple perfect-fluid form. The presence of $P$ induces momentum anisotropy, which is an expected result, since polarization defines a  priviliged direction in space. Such systems can be analyzed in the future with the methods worked out in the context of anisotropic hydrodynamics~\CITn{Florkowski:2010cf,Martinez:2010sc}.

\subsection{Spin tensor}
\label{sec:class-spin-tensor}
\medskip

The spin tensor is defined as an expectation value of the internal angular momentum tensor,
\bea
\hspace{-0.75cm}S^{\lambda, \mu\nu}_{\rm eq} &=& \int dP   \int dS \, \, p^\lambda \, s^{\mu \nu} 
\left[f^+_{\rm eq}(x,p,s) + f^-_{\rm eq}(x,p,s) \right] \nn \\
\hspace{-0.75cm}&=& 2 \cosh(\xi) \int dP \, p^\lambda \, \exp\LB - p \cdot \beta \RB
\int dS \, s^{\mu \nu} \, \exp\LB  \f{1}{2}  \omega_{\alpha \beta} s^{\alpha\beta} \RB.
\label{eq:Seq-sp01}
\eea
It is interesting first to check the leading-order approximation in $\omega$ for $S^{\lambda, \mu\nu}_{\rm eq}$. Expanding the exponential function we find
\bea
&&\hspace{-0.5cm}  \int dS \, s^{\mu \nu}  \, \exp\LB  \f{1}{2}  \omega_{\alpha \beta} s^{\alpha\beta}\RB =
\int dS \, s^{\mu \nu}  \, \left[1 +   \f{1}{2}  \omega_{\alpha \beta} s^{\alpha\beta}\right] 
\nn \\
&=&   \f{1}{2}  \omega_{\alpha \beta} \int dS \, s^{\mu \nu}  s^{\alpha\beta} =
\f{\omega_{\alpha \beta} }{2m^2}  \epsUmnrs \epsUabgd p_\rho p_\gamma  \int dS \,   s_\delta s_\sigma.
\label{eq:spinint3}
\eea
The last integral in the expression above is a function of momentum and, as a symmetric tensor, should have the following decomposition
\bea
\int dS \,   s_\delta s_\sigma = a\, g_{\delta \sigma} + b\, p_\delta p_\sigma,
\eea
where $a$ and $b$ are scalar functions obtained by the contractions:
\bea
\int dS \,   (p \cdot s)^2 &=& a\, m^2 + b\, m^4 = 0, \nn \\
\int dS \,   s^2  &=& 4 a + b\, m^2 = -2 \spin^2.
\eea
We find that $b = -a/m^2$ and $a = -(2/3) \spin^2$, which gives
\bea
\int dS \,   s_\delta s_\sigma = -\f{2}{3}\, \spin^2 \, \LB g_{\delta \sigma} -\f{ p_\delta p_\sigma}{m^2} \RB.
\label{eq:dSsdss}
\eea
Substituting \EQ{eq:dSsdss} into \EQ{eq:spinint3} and performing the contraction of the Levi-Civita tensors we find
\bea
&&\hspace{-0.0cm}  \int dS \, s^{\mu \nu}  \, \exp\LB  \f{1}{2 }  \omega_{\alpha \beta} s^{\alpha\beta}\RB =
\f{2}{3 m^2} \spin^2 \LB m^2 \omnU + 2 p^\alpha p^{[\mu} \omega^{\nu ]}_{\,\,\alpha} \RB. 
\label{eq:spinint4}
\eea
Using subsequently \EQ{eq:spinint4} in the definition of the spin tensor \EQn{eq:Seq-sp01} we find
\bea
S^{\lambda, \mu\nu}_{\rm eq} 
&=& \f{4}{3 m^2} \spin^2 \cosh(\xi) \int dP \, p^\lambda \, e^{ - p \cdot \beta }
\LB m^2 \omnU + 2 p^\alpha p^{[\mu} \omega^{\nu ]}_{\,\,\alpha} \RB. 
\label{eq:Seq-sp1}
\eea
It is striking to observe that with the value \EQn{eq:spin2} used for $\spin$ we reproduce the result \EQn{eq:Smunulambda_de_Groot2} in the case of small $\omega$.

In the case of arbitrary large $\omega$ or $P$, a similar calculation to that outlined above leads to the results
\bea
&&\hspace{-0.0cm}  \int dS \, s^{\mu \nu}  \, \exp\LB  \f{1}{2 }  \omega_{\alpha \beta} s^{\alpha\beta}\RB =
\f{\chi(P\spin)}{m^2} \LB m^2 \omnU + 2 p^\alpha p^{[\mu} \omega^{\nu ]}_{\,\,\alpha} \RB
\label{eq:spinint32}
\eea
and
\bea
S^{\lambda, \mu\nu}_{\rm eq} 
&=& \f{2 }{m^2} \cosh(\xi) \int dP \, \chi(P\spin)\, p^\lambda \, e^{ - p \cdot \beta }
\LB m^2 \omnU + 2 p^\alpha p^{[\mu} \omega^{\nu ]}_{\,\,\alpha} \RB,
\label{eq:Seq-sp2}
\eea
where the function $\chi(P\spin)$ is defined by the formula
\bea
\chi(P\spin) = \f{2 \left[P \spin \cosh(P \spin) - \sinh(P \spin)  \right]}{P^3 \spin}.
\label{eq:fchi}
\eea
For small values of $P$, we may use the approximation
\bea
\chi(P\spin) \approx \f{2\spin^2}{3} + \f{\spin^4 P^2}{15}.
\label{eq:fchi-app}
\eea
We thus see that in the leading order of the expansion in $P$, \EQ{eq:Seq-sp2} is reduced to \EQ{eq:Seq-sp1}.

\subsection{Pauli-Luba\'nski vector}
\label{sec:class-PL}
\medskip

We define the particle number current for both particles and antiparticles as
\bea
{\cal N}^\mu_{\rm eq} = \int dP   \int dS \, \, p^\mu \, \left[f^+_{\rm eq}(x,p,s)+f^-_{\rm eq}(x,p,s) \right].
\label{eq:calNeq-sp0}
\eea
Using this expression we obtain the momentum density of the total number of particles
\bea
E_p \f{d\Delta {\cal N}}{d^3p} = \f{\cosh(\xi)}{4\pi^3} \Delta\Sigma \cdot p \, e^{-p\cdot \beta} \int dS \,  \exp\LB  \f{1}{2}  \omega_{\alpha \beta} s^{\alpha\beta}\RB.
\eea
Similarly, the spin density defined by the spin tensor is
\bea
E_p \f{d\Delta {\cal S}^{\mu\nu}}{d^3p} = \f{\cosh(\xi)}{4\pi^3} \Delta\Sigma \cdot p \, e^{-p\cdot \beta}
\int dS \, s^{\mu \nu}  \, \exp\LB  \f{1}{2}  \omega_{\alpha \beta} s^{\alpha\beta}\RB
\eea
The phase-space density of the PL vector is then obtained as the ratio
\bea 
\pi_\mu(x,p) = -\f{1}{2} \epsLmnab
\f{\int dS \, s^{\nu \alpha}  \, \exp\LB  \f{1}{2}  \omega_{\rho \sigma} s^{\rho \sigma}\RB}{\int dS \,  \exp\LB  \f{1}{2}  \omega_{\rho \sigma} s^{\rho \sigma}\RB}
\f{p^\beta}{m}.
\eea 

Using \EQSTWO{eq:spinint2}{eq:spinint32} we find that the PL vector can be expressed by the simple expression
\bea
\pi_\mu = - \spin \, \f{{\tilde \omega_{\mu \beta}}}{P} \, \f{p^\beta}{m} \, \Lg(P  \spin),
\label{PL1}
\eea
where $\Lg(x)$ is the Langevin function defined by the formula
\bea 
\Lg(x) =  \coth(x) - \f{1}{x}.
\label{eq:Lg}
\eea 
It has the following expansions for large and small arguments: $\Lg \approx 1$ for $x \gg 1$ and $\Lg \approx \f{x}{3}$ for $x \ll 1$. In the particle rest frame we find
\bea 
\pi^0_* = 0, \quad \piv_* = - \spin \, \,\f{\Pv}{P} \,\, \Lg(P \spin) .
\eea 
Hence, in PRF the direction of the PL vector agrees with that of the polarization vector $\Pv$. For small and large $P$ we obtain two important results:
\bea 
\piv_* = -  \spin\,\f{\Pv}{P}, \quad
|\piv_*| = \spin = \sqrt{\f{3}{4}}, \quad
\text{if} \quad P \gg 1
\label{eq:PLlarge}
\eea 
and
\bea 
\piv_* = - \spin^2\,\f{\Pv}{3}, \quad
|\piv_*| = \spin^2\, \f{P}{3} = \f{P}{4}, \quad
\text{if} \quad P \ll 1.
\label{eq:PLsmall}
\eea 
Equation \EQn{eq:PLlarge} demonstrates that the normalization of the PL vector cannot exceed the value of $\spin$. On the other hand, \EQ{eq:PLsmall} shows that for small values of $P$ the classical treatment of spin reproduces the quantum mechanical result.

\subsection{Entropy conservation}
\label{sec:ent}
\medskip

Classical treatment of spin allows for explicit derivation of the entropy current conservation. For the latter we adopt the Boltzmann definition (see Sec. 40 in~\cite{Landau:1980mil} or Sec. 8.4 in \cite{Florkowski:2010zz})
\bea
H^\mu = - \int dP \int dS \, p^\mu 
\LSB 
f^+_{\rm eq} \LB \ln f^+_{\rm eq} -1 \RB  + 
f^-_{\rm eq} \LB \ln f^-_{\rm eq} -1 \RB \RSB.
\label{eq:H1}
\eea 
Using \EQ{eq:fpm-spin} and the conservation laws for energy, linear and angular momentum, and charge, we obtain
\bea
H^\mu = \beta_\alpha T^{\mu \alpha}_{\rm eq} -\f{1}{2} \omega_{\alpha\beta} S^{\mu, \alpha \beta}_{\rm eq}
-\xi N^\mu_{\rm eq} + {\cal N}^\mu_{\rm eq}
\label{eq:H2}
\eea 
and
\bea
\p_\mu H^\mu = \left( \p_\mu \beta_\alpha \right) T^{\mu \alpha}_{\rm eq} 
-\f{1}{2} \left( \p_\mu \omega_{\alpha\beta} \right) S^{\mu, \alpha \beta}_{\rm eq}
- \left(\p_\mu \xi \right)  N^\mu_{\rm eq} + \p_\mu {\cal N}^\mu_{\rm eq}.
\label{eq:H3}
\eea 
With the help of the relation
\bea
{\cal N}^\mu_{\rm eq} = \f{\cosh(\xi)}{\sinh(\xi)}  N^\mu_{\rm eq}
\label{eq:NcalN}
\eea 
that is valid for classical statistics, and the conservation of charge one can easily show that 
\bea 
\p_\mu H^\mu = 0.
\label{eq:entcon}
\eea
One can notice that the contributions to the entropy current~\EQn{eq:H2}, connected with the polarization tensor, start with quadratic terms in $\omega$. Hence, to account for the substantial polarization effects in a consistent matter, one should include also (at least) the second-order terms in the expressions for other thermodynamic quantities, such as the energy density or pressure. Nevertheless, if we restrict ourselves to linear terms in $\omega$, all thermodynamic quantities (including entropy) become independent of $\omega$, while the conservation of the angular momentum determines the polarization evolution in a given hydrodynamic background. This is important from the practical point of view, since the latter can be obtained from any hydrodynamic code without polarization and the spin effects can be studied on top of this.  
\section{Modeling of the space-time spin polarization in heavy-ion collisions}
\label{sec:appl}

In Sections \ref{sect:pfspinkin} and  \ref{sec:slmnGLW} we have shown that the most natural and self-consistent framework of relativistic perfect fluid hydrodynamics for spin-polarized systems can be formulated if one uses the GLW forms of the energy-momentum and spin tensors. Moreover, we have argued in Section \ref{sec:P3V} that the magnitude of the polarization three-vectors obtained with the spin density matrices is, in general, unbound from above, which restricts the applicability range of this framework to the case of small values of the spin chemical potential. Following these arguments, in this section we present a general scheme for construction of the GLW evolution equations of the perfect-fluid hydrodynamics with spin,  restricting ourselves to the case when spin polarization tensor is small. As a matter of fact, the latter choice represents a good approximation as it is supported by the experimental measurements which observe the global particle polarization on the order of a fraction of one percent~\cite{STAR:2017ckg,Adam:2018ivw}.
%
\subsection{Perfect fluid hydrodynamics with leading-order spin treatment}
\label{sec:LOpfhydro}

Performing the expansion of thermodynamic quantities in powers of the polarization tensor one can notice that the corrections start with quadratic terms.  Hence, in the leading (i.e. linear) order, the contributions from $\omega^{\a\b}$ to the charge density, energy density and pressure in Eqs.~(\ref{nden}), (\ref{enden}) and (\ref{prs})  may be neglected, 
\bea 
n &=& 4 \, \sinh(\xi)\, \nU(T), \label{n0small} \\
\varepsilon &=& 4 \, \cosh(\xi) \, \eU(T),\label{e0small}\\ 
P &=& 4 \, \cosh(\xi) \, \PU(T). \label{P0small} 
\eea
Here we have assumed that the  equation of state of the system is given by that of an ideal relativistic gas of massive Boltzmann particles and antiparticles with spin one half. In this case, using Eqs.~(\ref{nden0}), (\ref{enden0}) and (\ref{prs0}) in \EQSM{n0small} {P0small} and carrying out the momentum averaging one obtains the well-known results \cite{Florkowski:2010zz}
\beq
\nU(T) &=&  \f{ T^3}{2\pi^2}\,  \hat{m}^2 K_2\left( \hat{m}\right), \label{n0c}\\
\eU(T) &=& \f{ T^4 }{2\pi^2}  \, \hat{m} ^2
 \Big[ 3 K_{2}\left( \hat{m} \right) + \hat{m}  K_{1} \left( \hat{m}  \right) \Big],  \label{e0c}\\
\PU(T) &=& T \, \nU(T)  . \label{P0c}
\eeq
where $\hat{m}\equiv m/T$ is the ratio of the particle mass and temperature  and  $K_n\left( \hat{m}\right)$ denote the modified Bessel functions of the second kind.   

From the formulas (\ref{n0small})--(\ref{P0c}) one  can notice that for small polarization the dynamics of spin degrees of freedom decouples from the conservation laws for charge, energy, and linear momentum. In this case, these conservation laws yield ordinary evolution equations of perfect-fluid hydrodynamics for spinless particles. Using Eq.~(\ref{Nmu}) in the charge conservation \EQn{eq:Ncon} one gets 
\beq
u^{\a}\p_{\a}n+n\p_{\a}u^{\a}=0,
\label{eq:ncons}
\eeq
while using Eq.~(\ref{Tmn}) in the energy-momentum conservation  \EQn{eq:Tcon} and contracting with $u_\b$ and $\Delta^\mu_\b$  gives
\beq
u^{\a}\p_{\a}\varepsilon+(\varepsilon+P)\p_{\a}u^{\a}=0,\label{eq:encons}
\eeq
and
\beq
 (\varepsilon+P)u^{\a}\p_{\a}u^{\mu}-\Delta^{\mu\b}\p_{\b}P=0,\label{eq:momcons}
\eeq
respectively. Equations (\ref{eq:ncons})--(\ref{eq:momcons}) provide, in general, five independent partial differential equations for five independent hydrodynamic variables, $T$, $\xi$ and $u^\mu$ (note that $\varepsilon$ and $P$ are functions of $T$ and $\xi$, defined by the equation of state). The space-time evolution of $T$, $\xi$ and $u^\mu$ defines a thermal background for the dynamics of spin degrees of freedom. The latter is determined by the spin tensor conservation \EQn{eq:SGLWcon}.  

 Using the fact that the spin polarization tensor $\omega_{\mu\nu}$ is antisymmetric, one can introduce a decomposition of $\omega_{\mu\nu}$ with respect to the flow four-vector $u^\mu$ in terms of two new four-vectors $\kappa^\mu$ and $\omega^\mu$ \cite{Florkowski:2017ruc}, 
\beq
\omega_{\mu\nu} = \kappa_\mu u_\nu - \kappa_\nu u_\mu + \epsilon_{\mu\nu\a\b} u^\a \omega^{\b}. \label{spinpol1}
\eeq
Since any part of the four-vectors $\kappa^{\mu}$ and $\omega^{\mu}$  parallel to  $u^{\mu}$ does not contribute to the right-hand side of Eq.~(\ref{spinpol1}) we can assume the following orthogonality conditions 
\beq
\kappa\cdot u = 0, \quad \omega \cdot u = 0  \label{ko_ortho}.
\eeq
The latter allow us to express four-vectors $\kappa_\mu$ and $\omega_\mu$  in terms of the spin polarization tensor
\beq
\kappa_\mu= \omega_{\mu\a} u^\a, \quad \omega_\mu = \half \epsilon_{\mu\a\b\g} \omega^{\a\b} u^\g. \label{eq:kappaomega}
\eeq
Using \EQ{SDeltaGLW}, the spin tensor \EQn{eq:Smunulambda_de_Groot2} may be put in the following convenient  form
\beq
S^{\a, \b\g}_\GLW 
= {\cal A}_1 u^\a \omega^{\b\g} 
+ {\cal A}_2 u^\a u^{[\b} \kappa^{\g]} + {\cal A}_3 (u^{[\b} \omega^{\g]\a} + g^{\a[\b} \kappa^{\g]}) ,
\label{SGLW2}
\eeq
where we have defined
\beq
{\cal A}_1 &=& {\cal C} \LR \nU -  {\cal B}_{(0)} \RR \label{A1} ,\\ 
{\cal A}_2 &=& {\cal C} \LR {\cal A}_{(0)} - 3{\cal B}_{(0)} \RR  \label{A2} , \\ 
{\cal A}_3 &=&  {\cal C}\, {\cal B}_{(0)}\label{A3},
\eeq
with $
{\cal C}= \hbar  \ch{\xi}$. 
Using~\EQ{SGLW2} on the left-hand side of the conservation equation for the spin tensor \EQn{eq:SGLWcon}, one finds the following tensor equation
\beq
  & & \omega^{\b\g}  \LR \dot {\cal A}_1 + {\cal A}_1  \theta \RR  +{\cal A}_1  \dot {\omega}^{\b\g} \nn\\
&&\,+
     u^{[\b}  \kappa^{\g]} \LR\dot{\cal A}_2+  {\cal A}_2 \theta \RR+ {\cal A}_2   \LR \dot {u}^{[\b}  \kappa^{\g]} + u^{[\b}  \dot {\kappa}^{\g]}\RR \nn\\
  &&\,+\,
u^{[\b} \omega^{\g]\a} \p_\a    {\cal A}_3 + {\cal A}_3 \p_\a (u^{[\b} \omega^{\g]\a}) \nn\\
&&\,+\,
\p^{[\b} ({\cal A}_3\kappa^{\g]}) = 0. \label{dSmnl}
\eeq
In the next Section we will analyse equations (\ref{eq:ncons})--(\ref{eq:momcons}) and (\ref{dSmnl}) in the special case of the one-dimensional boost-invariant expansion. 
\subsection{One-dimensional Bjorken expansion }
\label{sec:bjorkenflow}
%
In the following, we consider a system that undergoes boost-invariant and transversely homogeneous expansion, most often dubbed the Bjorken flow \CITn{Bjorken:1982qr}. In such a situation, it is convenient to introduce the following local four-vector basis:
\beq
u^\a &=& \frac{1}{\tau}\LR t,0,0,z \RR = \LR \ch\eta, 0,0, \sh\eta \RR, \nn \\
X^\a &=& \LR 0, 1,0, 0 \RR,\nn\\
Y^\a &=& \LR 0, 0,1, 0 \RR, \nn\\
Z^\a &=& \frac{1}{\tau}\LR z,0,0,t \RR = \LR \sh\eta, 0,0, \ch\eta \RR,
\label{BIbasis}
\eeq
where $\tau = \sqrt{t^2-z^2}$ is the longitudinal proper time and $\eta =  \ln\LS(t+z)/(t-z)\RS/2$ is the space-time rapidity.  The four-vector $u^\a$ is identified with the flow vector of the fluid element, and in the local rest frame it reads $u^\a = (1,0,0,0)$.  The basis vectors \EQn{BIbasis} satisfy the following orthogonality and normalization conditions
\begin{eqnarray}
 u \cdot u &=& 1,\label{UU}\\
X \cdot X &=&   Y \cdot Y \,\,=\,\, Z \cdot Z \,\,=\,\, -1, \\ \label{XXYYZZ}
X \cdot u  \,\,&=& Y \cdot u\,\, \,\,=\,\, Z \cdot u \,\,=\,\, 0,  \\ \label{XYZU}
X \cdot Y  &=&  Y \cdot Z \,\,=\,\, Z \cdot X \,\,=\,\, 0.  \label{XYYZZX}
\end{eqnarray}
 By introducing the  derivatives with respect to $\tau$ and $\eta$, we can find the following expressions
\beq
\theta \equiv \p \cdot u = \frac{1}{\tau}, \quad&&  \,\dot{(\dots)} \equiv u \cdot \p = \p_\tau  , \\
\p \cdot X = 0, \,\quad&&\quad \quad\quad X \cdot \p = \p_x,\\
\p \cdot Y = 0, \,\quad&&\quad \quad\quad Y \cdot \p = \p_y,\\
\p \cdot Z = 0,\, \quad&&\quad \quad\quad Z \cdot \p = \frac{1}{\tau} \partial_{\eta}.
\eeq
With the formulas defined above, the charge conservation \rfn{eq:ncons} yields the equation
\beq
\dot{n}+\frac{n}{\tau}=0,\label{eq:charge}
\eeq
which has a simple scaling solution $n = n_0 \tau_0/\tau$, where $n_0\equiv n(\tau_0)$ is the charge density at the initial time $\tau_0$. On the other hand, the energy conservation equation \rfn{eq:encons} takes the form
\beq
\dot{\varepsilon}+\frac{(\varepsilon+P)}{\tau}=0,\label{eq:en}
\eeq
which (after employing thermodynamic relations) can be shown to be equivalent to the entropy conservation. In the case of one dimensional boost-invariant flow the remaining Euler equation \EQn{eq:momcons} is satisfied trivially. Equations (\ref{eq:charge})--(\ref{eq:en}) determine the evolution of temperature and chemical potential as functions of proper time, which gives the hydrodynamic background on top of which one has to solve \EQS{dSmnl} for the spin dynamics.
  
Using the basis \rfn{BIbasis} and employing conditions~\rfn{ko_ortho}, one can decompose four-vectors $\kappa^{\mu}$ and $\omega^{\mu}$ in the following way
\beq
\kappa^\a &=&  C_{\kappa X} X^\a + C_{\kappa Y} Y^\a + C_{\kappa Z} Z^\a, \label{eq:k_decom}\\
\omega^\a &=&  C_{\omega X} X^\a + C_{\omega Y} Y^\a + C_{\omega Z} Z^\a. \label{eq:o_decom}
\eeq
Using then Eqs.~\rfn{eq:k_decom} and \rfn{eq:o_decom} in \rf{spinpol1}, we get a boost-invariant form of the spin polarization tensor, 
\beq
\omega_{\mu\nu} &=& C_{\kappa Z} (Z_\mu U_\nu - Z_\nu U_\mu) \label{eq:omegamunu}   + C_{\kappa X} (X_\mu U_\nu - X_\nu U_\mu)    + C_{\kappa Y} (Y_\mu U_\nu - Y_\nu U_\mu) \nonumber \\
&& + \, \epsilon_{\mu\nu\alpha\beta} U^\alpha (C_{\omega Z} Z^\beta + C_{\omega X} X^\beta + C_{\omega Y} Y^\beta), \nn
\eeq
which can be used to get the boost-invariant form of the spin tensor, $S^{\a , \b\g}_{\GLW}$.
The boost invariance implies that all scalar functions, including coefficients  ${C}_{\kappa X}$, ${C}_{\kappa Y}$, ${C}_{\kappa Z}$, ${C}_{\omega X}$, ${C}_{\omega Y}$, and ${C}_{\omega Z}$ in \EQSM{eq:k_decom}{eq:o_decom}, may depend solely on the longitudinal proper time. Instead of studying the tensor equation \EQn{dSmnl},  it is convenient to consider its projections along the basis vectors defined by \rfn{BIbasis}. In particular, contracting \EQ{dSmnl} with  $U_\b X_\g$, $U_\b Y_\g$, $U_\b Z_\g$,  $Y_\b Z_\g$, $X_\b Z_\g$ and $X_\b Y_\g$ one gets the following evolution equations
\begin{equation}
{\rm diag}\LR
\cal{L}, \cal{L}, \cal{L}, \cal{P}, \cal{P}, \cal{P}\RR \,\,
\Dot{\Cv} ={\rm diag}\LR
{\cal{Q}}_1, {\cal{Q}}_1, {\cal{Q}}_2, {\cal{R}}_1, {\cal{R}}_1, {\cal{R}}_2 \RR\,\,
\Cv, \label{cs}
\end{equation} 
respectively, where $\Cv = \LR \Cv_\kappa, \Cv_\omega \RR$ is a six-component vector built from the $C$ coefficients which we earlier group into two three-vectors,
\beq 
\Cv_\kappa &=& (C_{\kappa X}, C_{\kappa Y}, C_{\kappa Z}), 
\label{Ckappa} \\
\Cv_\omega &=& (C_{\omega X}, C_{\omega Y}, C_{\omega Z}),
\label{Comega}
\eeq
moreover,
\beq
{\cal L}(\tau)&=&{\cal A}_1-\frac{1}{2}{\cal A}_2-{\cal A}_3,\nn\\
{\cal P}(\tau)&=&{\cal A}_1,\nn\\
{\cal{Q}}_1(\tau)&=&-\left[\dot{{\cal L}}+\frac{1}{\tau}\left( {\cal L}+ \frac{1}{2}{\cal A}_3\right)\right],\nn\\
 {\cal{Q}}_2(\tau)&=&-\left(\dot{{\cal L}}+\frac{{\cal L}}{\tau}   \right),\nn\\
  {\cal{R}}_1(\tau)&=&-\left[\Dot{\cal P}+\frac{1}{\tau}\left({\cal P} -\frac{1}{2} {\cal A}_3 \right) \right],\nn\\
 {\cal{R}}_2(\tau)&=&-\left(\Dot{{\cal P}} +\frac{{\cal P}}{\tau}\right).
 \label{LPQR}
 \eeq
 One can conclude from \EQ{cs} that each coefficient $C$ evolves independently. In addition, due to rotational invariance in the transverse plane, one can notice the coefficients $C_{\kappa X}$ and $C_{\kappa Y}$ as well as $C_{\omega X}$ and $C_{\omega Y}$ fulfill the same equations of motion.

\subsection{Total angular momentum of a boost-invariant cylinder}
\label{sec:fc}
\begin{figure} 
\centering
\includegraphics[width=0.45\textwidth]{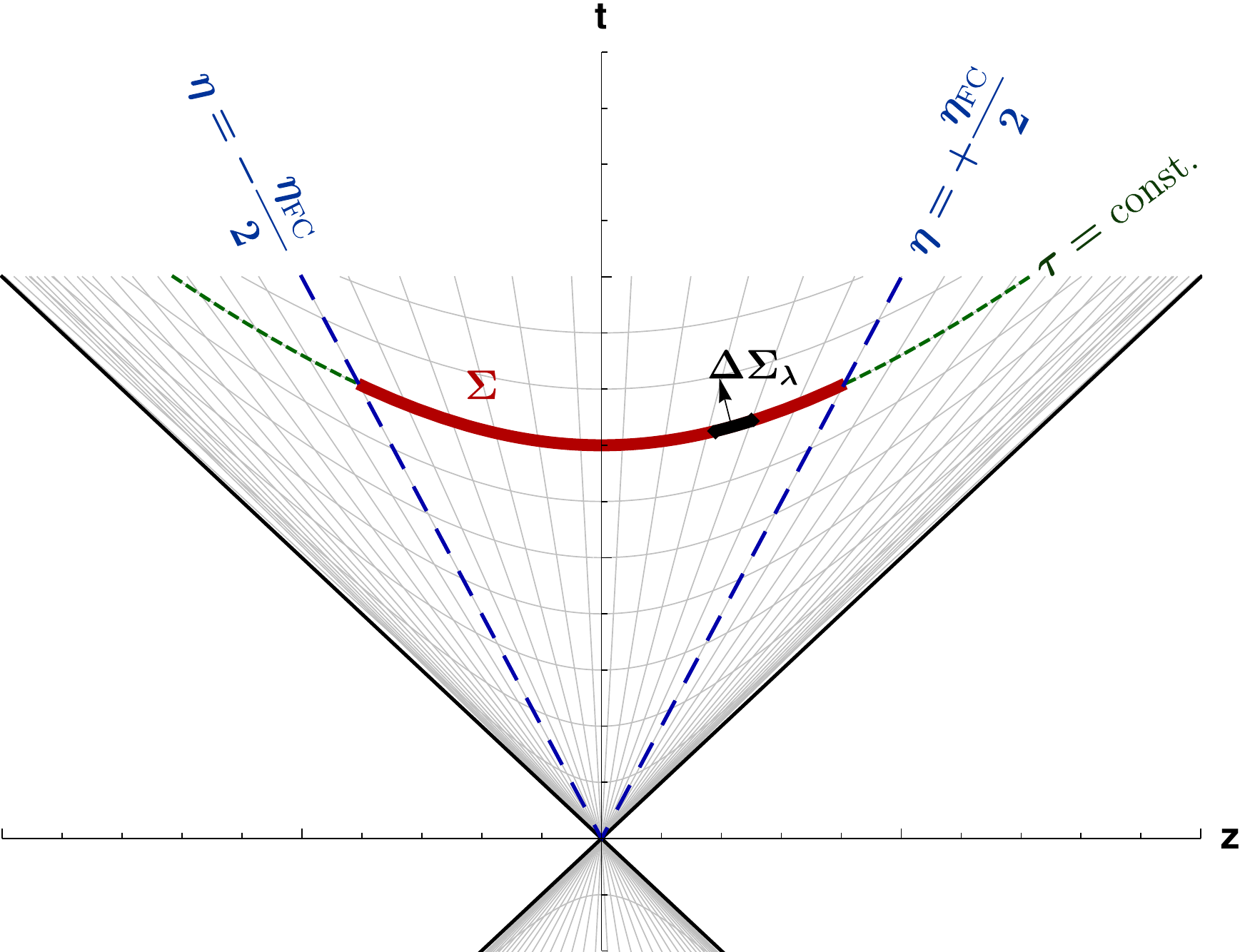}
\caption{(Color online) The hypersurface of the boost-invariant fire-cylinder.}
\label{fig:fc}
\end{figure}

%
In order to gain better physical interpretation of the $C$ coefficients, we will consider a boost-invariant fire-cylinder (FC) defined by the conditions

\beq
\tau={\rm const}, \qquad -\f{\eta_{\rm FC}}{2} \leq \eta \leq +\f{\eta_{\rm FC}}{2}, \qquad \sqrt{x^2+y^2} \leq R,
 \label{FCcond} 
\eeq
see Fig.~\ref{fig:fc}. A small space-time element of the fire-cylinder, $\Delta \Sigma _{\lambda }$, is given by the formula
\beq
\Delta \Sigma _{\lambda } &=& u_{\lambda }\, dx dy\, \tau  d\eta. \label{sig} 
\eeq
The spin part of the total angular momentum contained in the fire-cylinder is
\beq
S^{\mu\nu}_{\rm FC} &=& \int
\Delta \Sigma _{\lambda } S^{\lambda,\mu\nu}_{\rm GLW} =
\int
dx dy \,\tau  d\eta \, u_{\lambda } S^{\lambda,\mu\nu}_{\rm GLW} \nn \\ &=& A_\perp \tau \int\limits_{-\eta_{\rm FC}/2}^{+\eta_{\rm FC}/2} d\eta \, u_{\lambda } S^{\lambda,\mu\nu}_{\rm GLW} ,
\label{SAM}
\eeq
 with $A_\perp = \pi R^2$. Employing Eqs.~\rfn{SGLW2},  \rfn{eq:k_decom},  \rfn{eq:o_decom} and \rfn{sig},  \rf{SAM} can be rewritten the following form
\beq
S^{\mu\nu}_{\rm FC}&=& A_\perp  \tau \int\limits_{-\eta_{FC}/2}^{+\eta_{\rm FC}/2} d\eta   \left[ \vphantom{\int}  \right. {\cal A}_\kappa \big[ C_{\kappa X} \left(u^{\nu } X^{\mu }-u^{\mu } X^{\nu }\right)\nn\\&&+C_{\kappa Y} \left(u^{\nu } Y^{\mu }-u^{\mu } Y^{\nu }\right)+C_{\kappa Z} \left(u^{\nu } Z^{\mu }-u^{\mu } Z^{\nu }\right)\big] \nn\\&&+{\cal A}_1 \epsilon ^{\mu \nu \delta \chi } u_{\delta } \left( C_{\omega X} X_{\chi }+C_{\omega Y} Y_{\chi }+C_{\omega Z} Z_{\chi } \right) \left. \vphantom{\int}  \right], \nn \\
\label{SAMBI}
\eeq
where ${\cal A}_\kappa \equiv {\cal A}_1+ ({\cal A}_2+2{\cal A}_3)/2$. Using \rf{BIbasis} in  \rf{SAMBI} one can find that $S^{\mu\nu}_{\rm FC}$ is given by the following antisymmetric matrix
\begin{equation}
S^{\mu\nu}_{\rm FC} = -S^{\nu\mu}_{\rm FC} =
\begin{bmatrix}
\vspace{0.2cm}
0 & S^{01}_{\rm FC}  & S^{02}_{\rm FC} & S^{03}_{\rm FC} \\ \vspace{0.2cm}
-S^{01}_{\rm FC} & 0 & S^{12}_{\rm FC} & S^{13}_{\rm FC} \\ \vspace{0.2cm}
-S^{02}_{\rm FC} & -S^{12}_{\rm FC} & 0 & S^{23}_{\rm FC} \\ \vspace{0.1cm}
-S^{03}_{\rm FC} & -S^{13}_{\rm FC}  & -S^{23}_{\rm FC} & 0 \\
\end{bmatrix}, \label{LT}
\end{equation}
where 
\beq
S_{01}^{\rm FC}&=&2 A_\perp  \tau  \,{\cal A}_\kappa C_{\kappa X} \sinh\LR\f{\eta_{\rm FC}}{2}\RR, \nn\\
S_{02}^{\rm FC}&=&2A_\perp  \tau  \,{\cal A}_\kappa C_{\kappa Y} \sinh\LR\f{\eta_{\rm FC}}{2}\RR, \nn\\
S_{03}^{\rm FC}&=&A_\perp  \tau \, {\cal A}_\kappa C_{\kappa Z} \, \eta_{\rm FC}, \nn\\
S_{23}^{\rm FC}&=&-2 A_\perp  \tau \,{\cal A}_1 C_{\omega X} \sinh\LR\f{\eta_{\rm FC}}{2}\RR,  \nn\\
S_{13}^{\rm FC}&=&2A_\perp  \tau \,{\cal A}_1  C_{\omega Y} \sinh\LR\f{\eta_{\rm FC}}{2}\RR, \nn \\
S_{12}^{\rm FC}&=&-A_\perp \tau   \,{\cal A}_1 C_{\omega Z} \, \eta_{\rm FC}.
\label{Smunucoef}
\eeq
The above result clearly shows that the coefficients $C$ are related to different components of the total spin angular momentum of the boost-invariant fire-cylinder.

At this place it is also interesting to discuss the orbital contribution to the total angular momentum of the fire-cylinder. It is given by the expression
\beq
L^{\mu\nu}_{\rm FC} =\int
\Delta \Sigma _{\lambda } L^{\lambda,\mu\nu} =\int
\Delta \Sigma _{\lambda } \lt(x^{\mu}T^{\lambda\nu}_{\rm GLW}
-x^{\nu}T^{\lambda\mu}_{\rm GLW} \rt).
\nn\\\label{OAM}
\eeq
Using Eqs.~\rfn{Tmn} and \rfn{sig} in \rf{OAM} we can write,
\beq
L^{\mu\nu}_{\rm FC} =\int
dx dy\, \tau  d\eta \,\varepsilon \lt(x^{\mu}u^{\nu}-x^{\nu}u^{\mu}\rt).
\eeq
Substituting $u^{\mu}$ from \rf{BIbasis} into this equation, one can easily show that for our system
\beq
L^{\mu\nu}_{\rm FC} =0.
\eeq
Thus, the only finite contribution to the total angular momentum comes from the spin part.

\subsection{Numerical solutions}
\label{sec:hydro}

In this section we present numerical solutions of Eqs.~\rfn{eq:charge}, \rfn{eq:en} and \rfn{cs}. In the first step, using Eqs.~\rfn{eq:charge} and \rfn{eq:en} we determine the  proper-time dependence of the temperature $T$ and the chemical potential $\mu$, which allow us to calculate the functions ${\cal L}$, ${\cal P}$, ${\cal Q}$ and ${\cal R}$ appearing in  Eq.~\rfn{cs}. In the second step we solve  \rfn{cs} for the $C$ coefficients  which constitute the polarization tensor $\omega^{\mu\nu}$.

In order to investigate the physical situations similar to that studied in the heavy-ion collision experiments at low and intermediate energies, we consider a system with large initial baryon chemical potential $\mu_0=800$~MeV and low initial temperature $T_0=155$~MeV. We assume the system to be composed of the $\Lambda$-hyperons with the mass $m~=~1116$~MeV. We initialize hydrodynamic evolution at the proper time $\tau_0=1$~fm and  finish at the proper time $\tau_f=$~10~fm.

\begin{figure}
\centering
\includegraphics[width=0.6\textwidth]{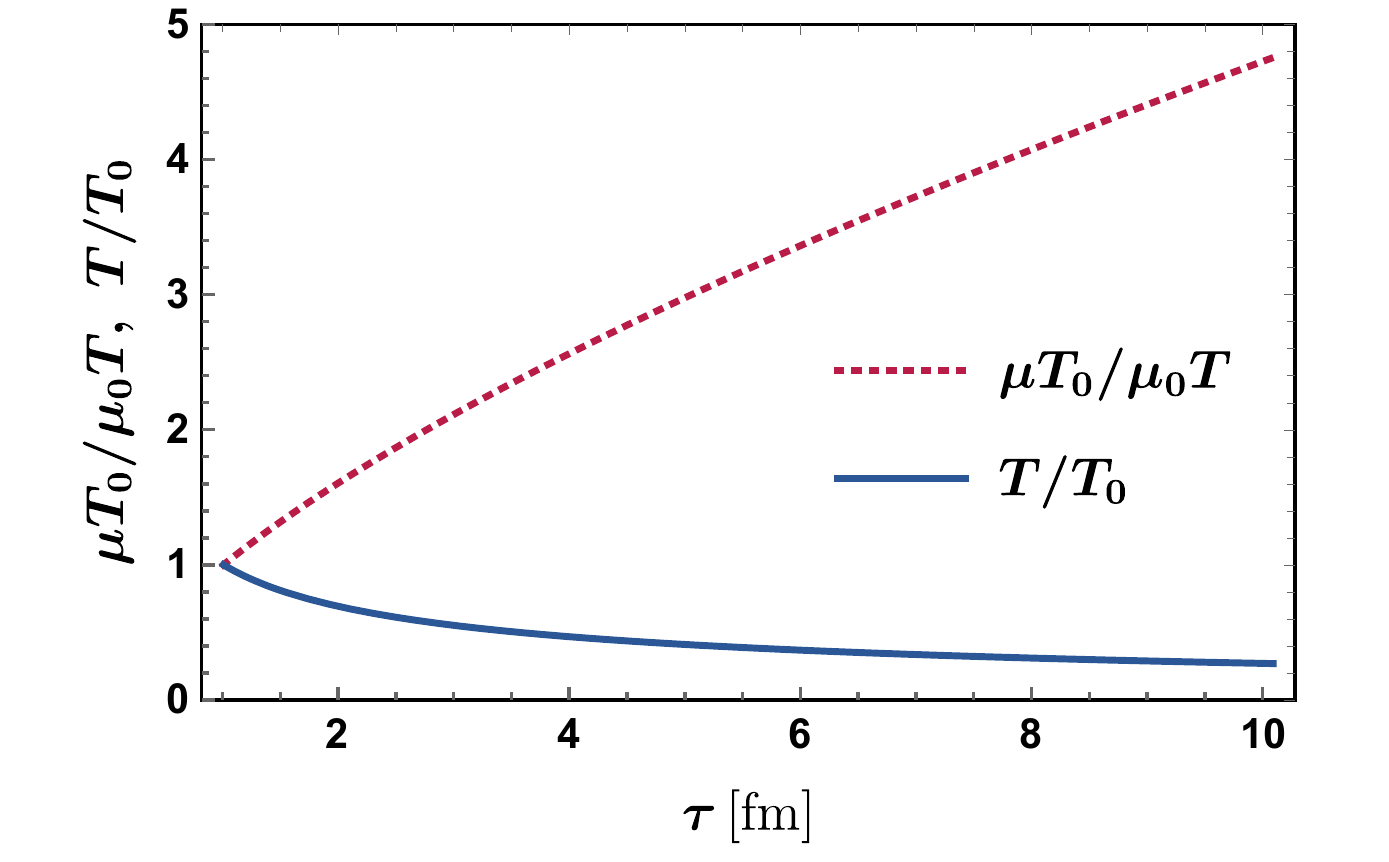}
\caption{Proper-time dependence of temperature $T$  scaled by its initial value $T_0$ (solid line) and the ratio of baryon chemical potential and temperature $\mu/T$  scaled by the initial ratio $\mu_0/T_0$ (dotted line) for a boost-invariant transversally homogenous expansion defined by Eqs.~\rfn{eq:charge} and \rfn{eq:en}.}
\label{fig:Tmu}
\end{figure}

\begin{figure}
\centering
\includegraphics[width=0.6\textwidth]{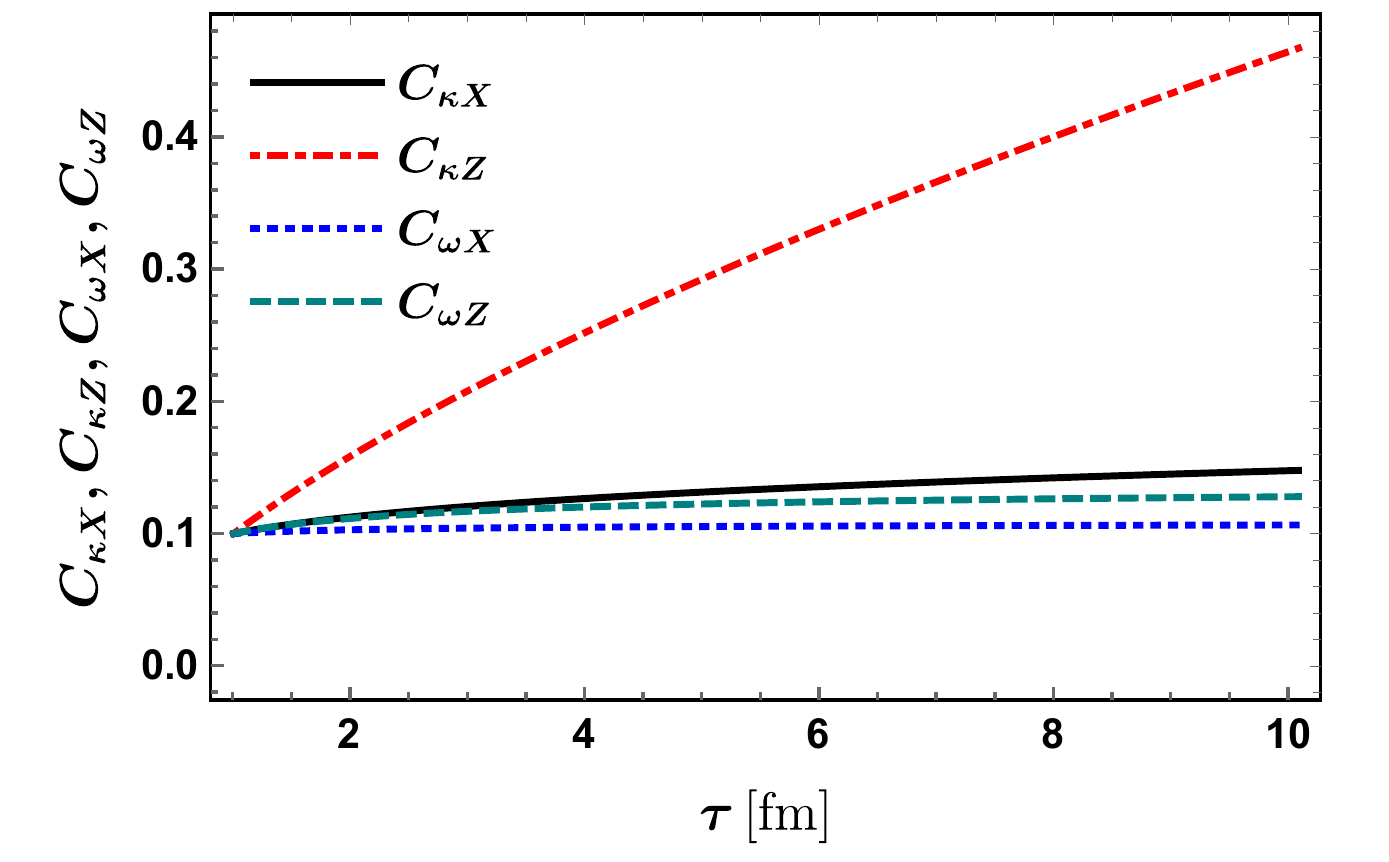}
\caption{Proper-time dependence of the coefficients $C_{\kappa X}$, $C_{\kappa Z}$, $C_{\omega X}$ and $C_{\omega Z}$ determined from Eq.~\rfn{cs}. Since the coefficients $C_{\kappa Y}$ and $C_{\omega Y}$ satisfy the same evolution equations as the coefficients $C_{\kappa X}$ and $C_{\omega X}$ we do not show the former here. }
\label{fig:c_coef}
\end{figure}
In Fig.~\ref{fig:Tmu} we show the proper-time dependence of the temperature and the baryon chemical potential determined from Eqs.~\rfn{eq:charge} and \rfn{eq:en}. Since, in this case the hydrodynamic evolution decouples completely from the dynamics of the spin, Eqs.~\rfn{eq:charge} and \rfn{eq:en} represent the state-of-the-art perfect-fluid hydrodynamic equations for charge conserving system. The spin degrees of freedom enter here through the trivial degeneracy factors in the thermodynamic quantities describing equation of state of the medium.  In consequence, in Fig.~\ref{fig:Tmu} we reproduce well established results with temperature decreasing and the ratio of the chemical potential to temperature increasing  with $\tau$. We note that in the case of massless particles the Bjorken expansion results in $\mu/T$ being constant in $\tau$ and the temperature decreasing as $T = T_0 (\tau_0/\tau)^{1/3}$.

The knowledge of the proper time dependence of the temperature and the chemical potential allows us to  determine  evolution of the components of spin polarization tensor as given by the $C$ coefficients using Eq.~\rfn{cs}. In order to compare the relative importance of the latter their initial values have been assumed to be the same.  In Fig. \ref{fig:c_coef} we present the proper-time dependence of the coefficients $C_{\kappa X}$, $C_{\kappa Z}$, $C_{\omega X}$ and $C_{\omega Z}$. Since the coefficients $C_{\kappa Y}$ and $C_{\omega Y}$ obey the same equations as $C_{\kappa X}$ and $C_{\omega X}$ we do not show the former here. From Fig. \ref{fig:c_coef} one can observe that all the coefficients have a  rather weak proper time dependence except for the coefficient $C_{\kappa Z}$, which increases by about $0.1$ within $1$ fm. Based on these results we may conclude that the inclusion of leading-order terms in polarization tensor in the hydrodynamic evolution is sufficient if the initial values of the coefficients $C$ are small and the evolution time is shorter than $\sim~10$ fm.    

\subsection{Spin polarization of particles at freeze-out}
%
In this section we use the proper-time evolution of the temperature, chemical potential and spin polarization tensor resulting from the hydrodynamic model to study the polarization of particles emitted at the freeze-out. For that purpose, we choose the same freeze-out hypersurface  as in the case of the boost-invariant fire-cylinder studied in Section \ref{sec:fc}. Subsequently, knowing form of the element of the freeze-out hypersurface $\Delta\Sigma_\mu$ we calculate the average PL vector of particles with momentum $p$. Finally,  perfoming canonical boost of the PL vector to the rest frame of the particles, we can determine their  average spin polarization. The latter quantity can be directly compared with the experimental data. 
\bigskip
\subsubsection{Average Pauli-Luba\'nski vector}
Using  Eq.~\rfn{eq:Smunulambda_de_Groot22} and keeping leading order terms in the spin polarization tensor the  phase-space density of the GLW spin tensor can be expressed by the following formula~\cite{Florkowski:2018ahw}
\beq
E_p \f{dS^{\lambda , \nu \a }_{\rm GLW}}{d^3p} &=&\frac{{\cosh}(\xi)}{(2\pi)^3 m^2} \, e^{-\beta \cdot p} p^{\lambda } \left(m^2\omega ^{\nu\a}+2 p^{\delta }p^{[\nu }\omega ^{\a ]}{}_{\delta } 
\right).  \label{eq:SGLW22} 
\eeq
Employing Eq.~\rfn{eq:SGLW22} in the definition of the phase-space density of the PL four-vector, see Eq.~\rfn{PL10},
\begin{equation}
E_p\frac{d\Delta \Pi _{\mu }(x,p)}{d^3 p}
=-\frac{1}{2}\epsilon _{\mu \nu \alpha \beta }\Delta 
\Sigma _{\lambda }E_p\frac{dS^{\lambda ,\nu \alpha }_{\rm GLW}(x,p)}
{d^3 p}\frac{p^{\beta }}{m},
\label{PL1}
\end{equation}
one may find the total (integrated over the freeze-out hypersurface) PL four-vector for particles with momentum $p$,
\begin{equation}
E_p\frac{d\Pi _{\mu }(p)}{d^3 p} = -\f{ \cosh(\xi)}{(2 \pi )^3 m}
\int
\Delta \Sigma _{\lambda } p^{\lambda } \,
e^{-\beta \cdot p} \,
\tilde{\omega }_{\mu \beta }p^{\beta }. \label{PDPLV}
\end{equation}
In the case of the Bjorken expansion considered here it is convenient to  introduce the following parametrization of the  four-momentum, 
\beq
p^\lambda &=& \left( m_T\ch y_p,p_x,p_y,m_T\sh y_p \right), \label{pl}
\eeq
where $m_T$ is its transverse mass and  $y_p$ its rapidity. Equation \rfn{pl} allows us to express the contraction of the dual polarization tensor and four-momentum in  \rf{PDPLV} in the following form
\begin{equation}
\tilde{\omega }_{\mu \beta }p^{\beta }=\left[
\begin{array}{c}
\phantom{-}\left(C_{\kappa X} p_y-C_{\kappa Y} p_x\right)\sinh (\eta)+\left(C_{\omega X} p_x+C_{\omega Y} p_y\right)\cosh (\eta)+C_{\omega Z} m_T\sinh \left(y_p\right)\\ \\
 \phantom{-}C_{\kappa Z} p_y-C_{\omega X} m_T \cosh \left(y_p-\eta \right)-C_{\kappa Y} m_T \sinh \left(y_p-\eta \right) \\ \\
 -C_{\kappa Z} p_x-C_{\omega Y} m_T \cosh \left(y_p-\eta \right)+C_{\kappa X} m_T\sinh \left(y_p-\eta \right) \\ \\
-\left(C_{\kappa X} p_y-C_{\kappa Y} p_x\right)\cosh(\eta )-\left(C_{\omega X} p_x+C_{\omega Y} p_y\right)\sinh(\eta)-C_{\omega Z} m_T\cosh\left(y_p\right) \\
\end{array}
\right]\,.\label{OP}
\end{equation}
Using the definition of the boost-invariant flow four-vector given in Eq.~\rfn{BIbasis} and form of the element of the freeze-out hypersurface in Eq.~\rfn{sig} with  \rf{pl} one gets
\begin{equation}
p^{\lambda }\beta_{\lambda}= \frac{m_T}{T}\cosh\left(y_p-\eta \right)\label{PU}   
\end{equation}
and
\begin{equation}
\Delta \Sigma _{\lambda }p^{\lambda }
= m_T\cosh\left(y_p-\eta \right)dx dy\, \tau d\eta, \label{SIGP}   
\end{equation}
respectively. 

Employing Eqs.~\rfn{OP} -\rfn{SIGP} in Eq.~\rfn{PDPLV} one may express the total PL vector by a combination of the modified Bessel functions 
\begin{equation}
E_p\frac{d\Pi _{\mu }(p)}{d^3 p}=C_1 K_{1}\left( \hat{m}_T \right)\left[\begin{array}{c}
-\chi\Big[ \left(C_{\kappa X} p_y-C_{\kappa Y} p_x\right)\sinh (y_p)+\left(C_{\omega X} p_x+C_{\omega Y} p_y\right)\cosh (y_p)\Big] -2 C_{\omega Z} m_T\sinh \left(y_p\right)  \\ \\
 - \big(2 C_{\kappa Z} p_y  -\chi C_{\omega X} m_T\big)\\ \\
\phantom{-} 2 C_{\kappa Z} p_x  +\chi C_{\omega Y} m_T \\ \\
\phantom{-} \chi\Big[\left(C_{\kappa X} p_y-C_{\kappa Y} p_x\right)\cosh (y_p)+\left(C_{\omega X} p_x+C_{\omega Y} p_y\right)\sinh (y_p)\Big] +2 C_{\omega Z} m_T\cosh \left(y_p\right) \\
\end{array}
\right],
\label{totPL}
\end{equation} 
where $\chi\left( \hat{m}_T \right)=\left( K_{0}\left( \hat{m}_T \right)+K_{2}\left( \hat{m}_T \right)\right)/K_{1}\left( \hat{m}_T \right)$, $\hat{m}_T=m_T/T$,  and 
\begin{equation}
C_1=\frac{A_\perp  \cosh(\xi) \tau m_T}{(2 \pi )^3 m}.
\end{equation} 
The average spin polarization per particle is defined by the ratio 
\beq
\langle\pi_{\mu}\rangle=\frac{E_p\frac{d\Pi _{\mu }(p)}{d^3 p}}{E_p\frac{d{\cal{N}}(p)}{d^3 p}},
\label{averagePL}
\eeq
where $E_p\frac{d{\cal{N}}(p)}{d^3 p}$  denotes the momentum density of all particles (i.e., particles and antiparticles). The latter is expressed by the formula
\beq
E_p\frac{d{\cal{N}}(p)}{d^3 p}&=&
\f{4 \cosh(\xi)}{(2 \pi )^3}
\int
\Delta \Sigma _{\lambda } p^{\lambda } 
\,
e^{-\beta \cdot p} \,,
\label{densityofpart}
\eeq
see \rf{DcalN}.
The integration over space-time rapidity and transverse space coordinates  in \rf{densityofpart} yields
\beq
E_p\frac{d{\cal{N}}}{d^3 p}&=&8 m C_1 K_{1}\left( \hat{m}_T \right).
\eeq 
Due to the classical statistics considered herein,  the coefficient $C_1$ cancels out in ratio \rfn{averagePL}. It implies that $\langle\pi_{\mu}\rangle$ does not depend explicitly on the chemical potential of the medium.

The polarization vector in the local rest frame of the particle, $\langle\pi^{\star}_{\mu}\rangle$, can be obtained by perfomring the canonical boost \cite{Leader:2001gr}
\beq
\langle\pi^{\star}_{\mu}\rangle=-\frac{1}{8m }\left[\begin{array}{c}
0 \\ \\
\left(\frac{p_x\sh y_p }{m_T \ch y_p+m}\right)\left[\chi\left(C_{\kappa X} p_y-C_{\kappa Y} p_x\right)+2 C_{\omega Z} m_T  \right] +
  \frac{ \chi \,p_x \ch y_p  \left(C_{\omega X} p_x+C_{\omega Y} p_y\right)}{m_T \ch y_p+m}\!+\!2 C_{\kappa Z} p_y  \!-\!\chi C_{\omega X}{m}_T \\ \\
\left(\frac{p_y\sh y_p }{m_T \ch y_p+m}\right)\left[\chi\left(C_{\kappa X} p_y-C_{\kappa Y} p_x\right)+2 C_{\omega Z} m_T  \right] + \frac{\chi \,p_y \ch y_p  \left(C_{\omega X} p_x+C_{\omega Y} p_y\right)}{m_T \ch y_p+m}\!-\!2 C_{\kappa Z} p_x \!-\!\chi C_{\omega Y}{m}_T \\ \\
 -\left(\frac{m\ch y_p+m_T}{m_T \ch y_p+m}\right)\left[\chi\left(C_{\kappa X} p_y-C_{\kappa Y} p_x\right)+2 C_{\omega Z} m_T  \right]
-\frac{\chi \,m\,\sh y_p\left(C_{\omega X} p_x+C_{\omega Y} p_y\right)}{m_T \ch y_p+m} \nn\\
\end{array}
\right]. \\
\label{PLVPPLPRF}  
\eeq 
As expected, the time component of the polarization vector $\langle\pi^{\star}_{\mu}\rangle$ vanishes, since in the particle rest frame one should have $\langle\pi^{\star}_{\mu}\rangle p^\mu_\star = \langle\pi^{\star}_{0}\rangle m =0$.

In the context of the  spin polarization measurements of $\Lambda$ hyperons done recently it is instructive to consider special case of particles emitted at midrapidity where $y_p = 0$. Also, since the mass of the $\Lambda$ particle is much larger than the values of temperature used in the hydrodynamic simulations, $\hat{m}_T \gg 1$, one may employ the approximation  $\chi(\hat{m}_T) \approx 2$. 
In consequence, \rf{PLVPPLPRF} takes the following compact form
\beq
\langle\pi^{\star}_{\mu}\rangle&=&-\frac{1}{4m}\left[\begin{array}{c}
0 \\ \\
\frac{ p_x \left(C_{\omega X} p_x+C_{\omega Y} p_y\right)}{m_T+m}+ C_{\kappa Z} p_y - C_{\omega X}{m}_T \\ \\
\frac{p_y \left(C_{\omega X} p_x+C_{\omega Y} p_y\right)}{m_T+m}- C_{\kappa Z} p_x -  C_{\omega Y}{m}_T \\ \\
- \left(C_{\kappa X} p_y-C_{\kappa Y} p_x\right) -  C_{\omega Z} m_T 
 \\
\end{array}
\right]. 
\label{PLVPPLPRFapp1}
\eeq
Introducing the three-vector notation for the spatial part of the polarization four-vector, $\langle \piv^* \rangle = (\langle\pi^{\star1}\rangle, \langle\pi^{\star2}\rangle, \langle\pi^{\star3}\rangle) \equiv (\langle\pi^{\star}_x\rangle, \langle\pi^{\star}_y\rangle, \langle\pi^{\star}_z\rangle)$, and using definitions \rfn{Ckappa} and \rfn{Comega} one may put \rf{PLVPPLPRFapp1} in the following simple form
\beq
\langle \piv^* \rangle = -\frac{1}{4m} \left[
E_p \Cv_\omega - \pv \times \Cv_\kappa - \frac{\pv \cdot \Cv_\omega}{E_p + m} \pv
\right], \label{simpleP}
\eeq
where $\pv = (p_x, p_y, 0)$. From \rf{simpleP} one may deduce that for particles with small transverse momenta the polarization three-vector $\langle \piv^* \rangle$  is directly related to the coefficients $\Cv_\omega$. Furthermore, due to the different proper-time dependence of the different coefficients $C$, see Fig.~\ref{fig:c_coef},  the mean polarization three-vector will exhibit non trivial changes of both length and direction.
 
%
\subsubsection{Momentum dependence of polarization}
%
In this section, using the methodology put forward above we apply our  hydrodynamical model to describe the polarization of particles produced in noncentral relativistic heavy ion collisions. It is usually argued that in the collision process the two incoming nuclei deposit large orbital angular momentum in the midrapidity region which is perpendicular to the reaction, $x-z$, plane and negative with respect to the $y$ axis, see Fig.~\rfn{fig:coll}. It is expected that, due to total angular momentum conservation, soon after the collision certain fraction of the  initial orbital angular momentum is transferred to the spin degrees of freedom~\cite{Voloshin:2004ha}. In order to address this physical situation we assume that the initial spin angular momentum has the same direction as the total angular momentum of the colliding system. In the numerical simulations this is achieved by choosing  $C_{\omega Y}(\tau=\tau_0) > 0$ (see \rf{Smunucoef}) and setting all other $C$ coefficients to zero. Using the initial conditions $\mu_0=800$~MeV,  $T_0=155$~MeV, $\Cv_{\kappa, 0} = (0,0,0)$, and $\Cv_{\omega, 0} = (0,0.1,0)$ we numerically solve hydrodynamic equations  and extract  the values of the thermodynamic parameters and the coefficients $C$ at freeze-out. The latter are used in  \rf{PLVPPLPRF}  to determine  different components of the polarization three-vector as functions of the particle transverse momentum components $p_x$ and $p_y$. In order to address the actual experimental measurements, which are done mainly at midrapidity, we set  $y_p=0$ ($p_z=0$). 

\begin{figure*}
      \centering
		\subfigure[]{}\includegraphics[width=0.32\textwidth]{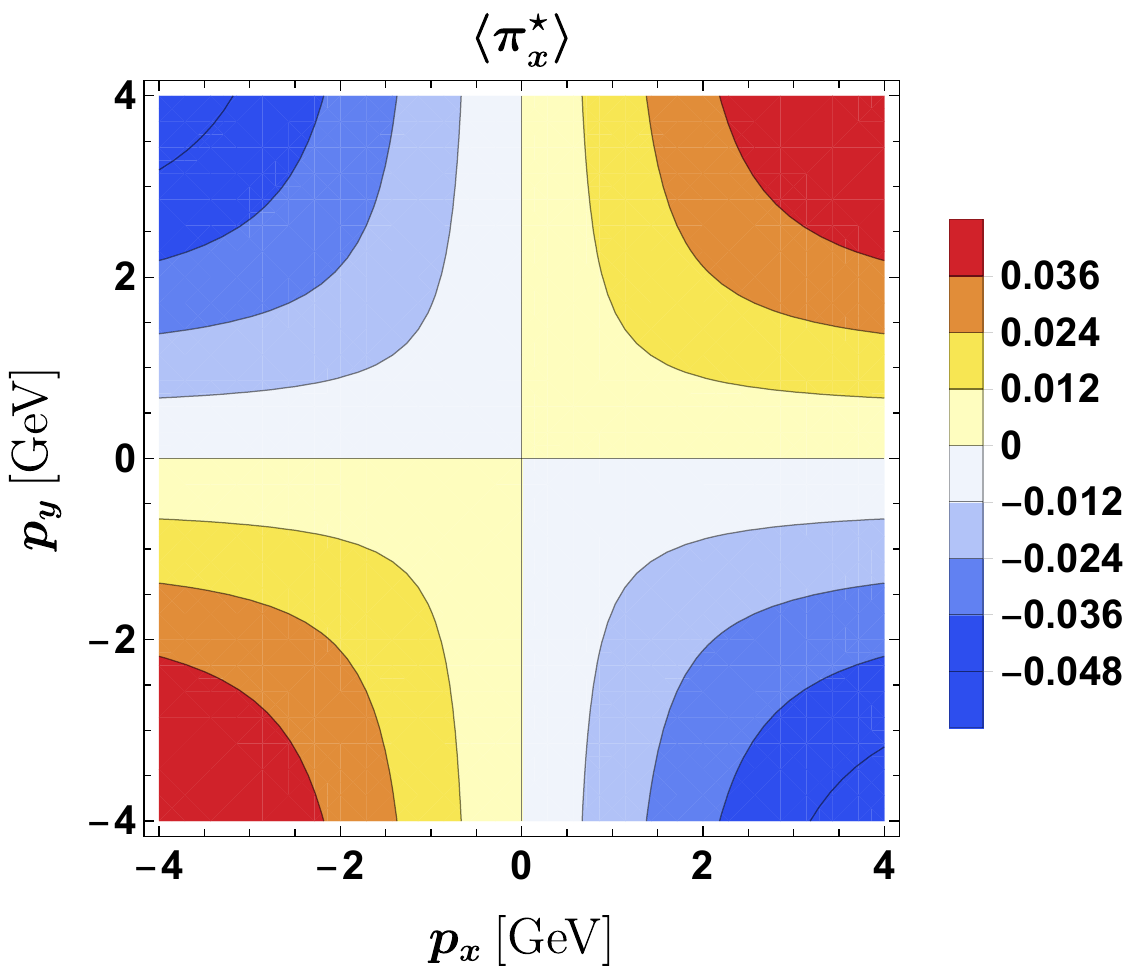}
		\subfigure[]{}\includegraphics[width=0.32\textwidth]{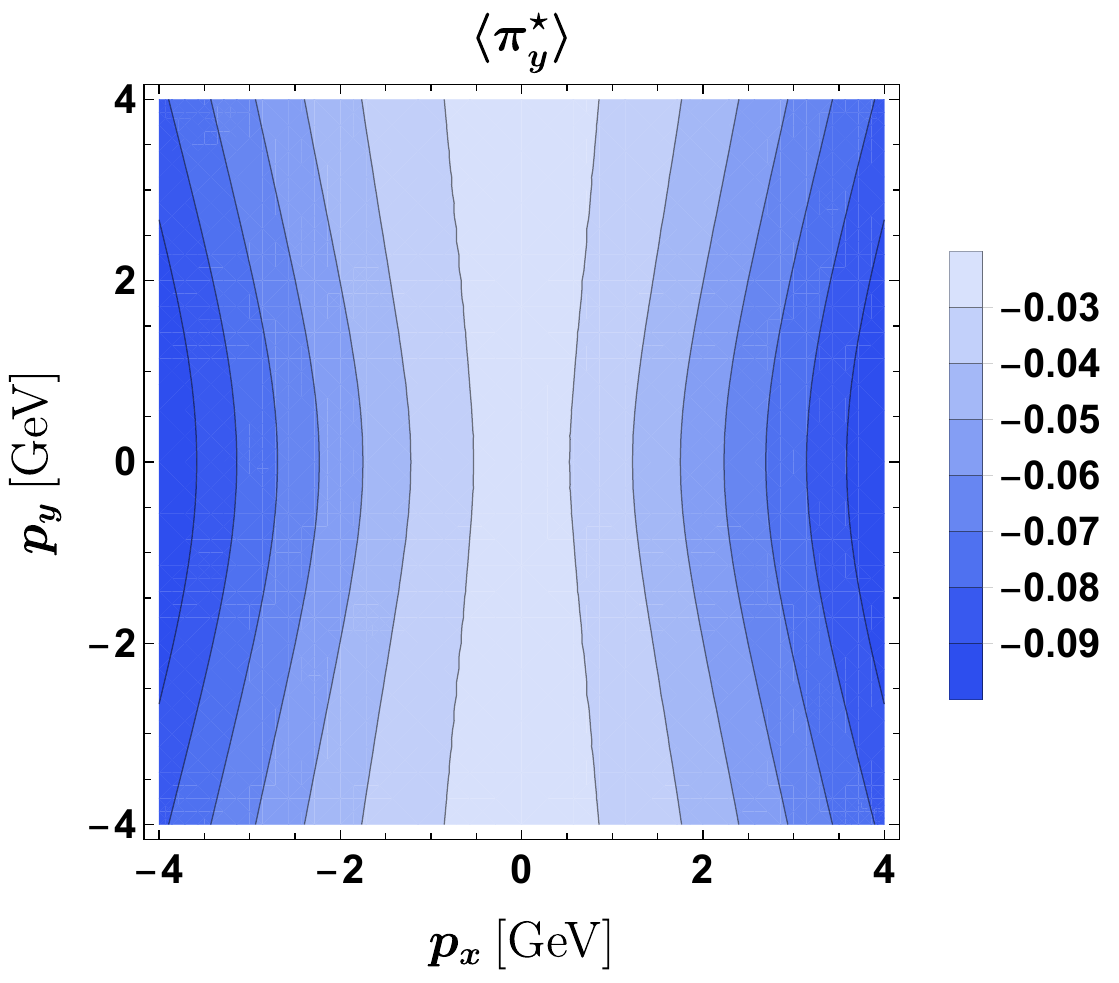}
		\subfigure[]{}\includegraphics[width=0.32\textwidth]{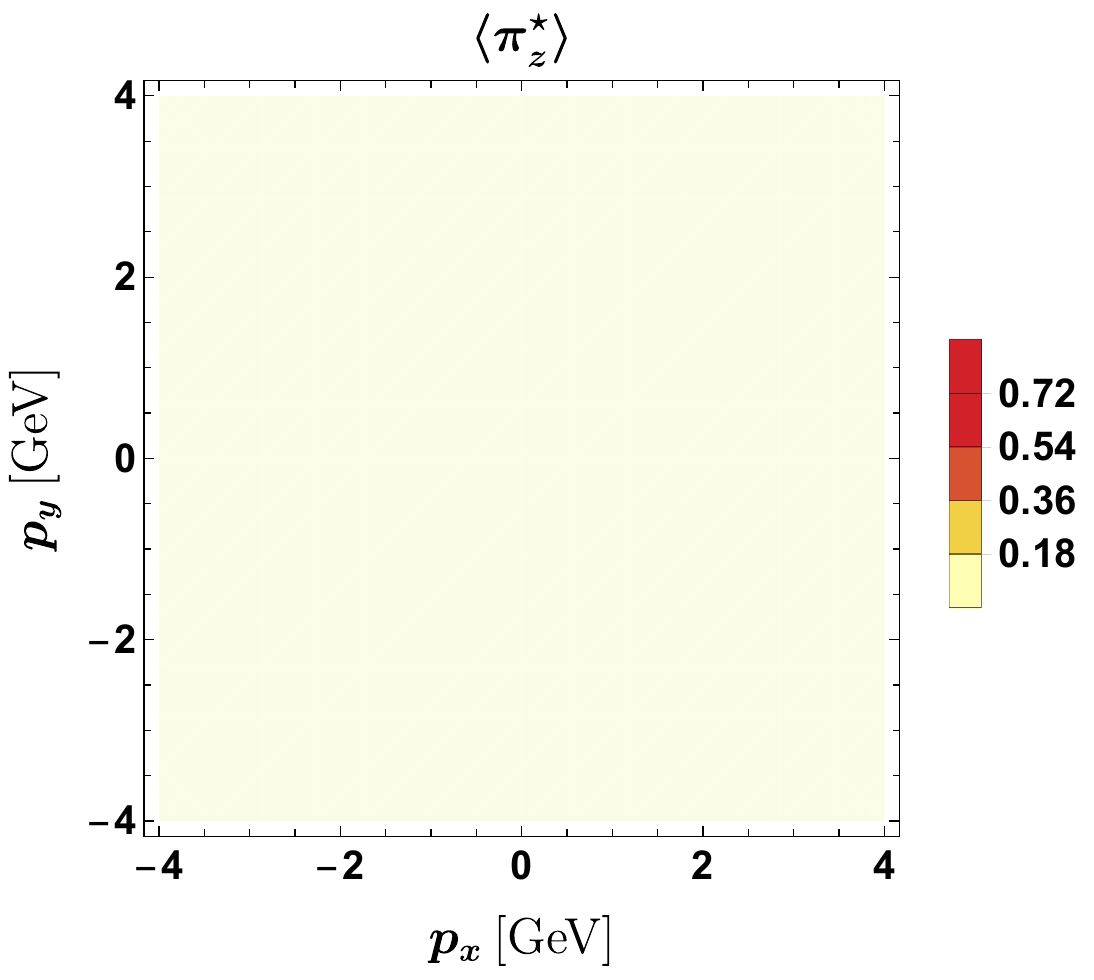}
		\caption{Components $\langle\pi^{\star}_x\rangle$ (left),  $ \langle\pi^{\star}_y\rangle$ (middle), and $ \langle\pi^{\star}_z\rangle$ (right) of PRF  mean polarization three-vector of $\Lambda$ hyperons. The results were obtained for $y_p=0$   at the freeze-out time $\tau_{f}=10$ fm with the initial conditions $\mu_0=800$~MeV,  $T_0=155$~MeV, $\Cv_{\kappa, 0} = (0,0,0)$, and $\Cv_{\omega, 0} = (0,0.1,0)$ set at $\tau_0 = 1$ fm.} 
	 \label{fig:polarization1}
\end{figure*}

The results of our calculations for the  three components of the polarization three-vector, $\langle\pi^{\star}_x\rangle$,  $ \langle\pi^{\star}_y\rangle$, and $ \langle\pi^{\star}_z\rangle$, as functions of the transverse momentum components $p_x$ and $p_y$ are shown in left, middle and right panels of Fig.~\ref{fig:polarization1}, respectively. As expected, the $ \langle\pi^{\star}_y\rangle$ component (see the middle panel of Fig.~\ref{fig:polarization1}) is negative, which reflects our choice of the direction of the initial spin angular momentum of the system. Since in our study we set $y_p=0$, the longitudinal component $\langle\pi^{\star}_z\rangle$ is vanishing completely (see the right panel). Consequently, the quadrupole structure observed in the experimental data is not reproduced. The latter result may be traced back to the symmetries imposed in our model. In particular, in the modeling we do not include  the elliptic deformation of the system in the transverse plane which is believed to be responsible for the non-trivial pattern seen in the data. However, one should stress here that the inclusion of the elliptic flow in the standard hydrodynamic models without spin assuming equality between thermal vorticity and spin polarization tensor is also not sufficient to describe the data. In fact, the most recent hydrodynamic-based results does not capture the correct sign of the latter. In the case of the component $\langle\pi^{\star}_x\rangle$ we observe a quadrupole structure (see the left panel) with signs changing in subsequent quadrants. It is interesting to observe that the sign different obtained within current study is oposite to the one obtained in other hydrodynamical calculations~\cite{Karpenko:2016jyx}.
   
 \begin{figure}
\centering
\includegraphics[width=0.45\textwidth]{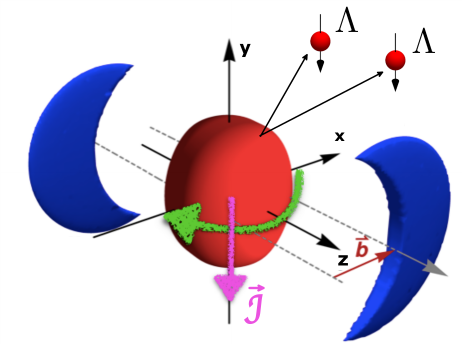}
\caption{The orientation of the initial orbital angular momentum deposited in the system in the inital stage of noncentral relativistic heavy-ion collision.}
\label{fig:coll}
\end{figure}%
 
Summarizing, the realistic modeling of the spin polarization dynamics in heavy ion collisions clearly requires full (3+1)-dimensional hydrodynamic models. Breaking the symmetries seems to be  crucial for addressing  the phenomena observed in experiment. The presented study outline the metodology which may be used to determine the spin polarization of particles from the hydrodynamical models with spin.

%
\section{Conclusions}
\label{sec:con}

We close the paper with the following series of comments:

\begin{itemize} 
	\item[{ 1.}] The arguments collected in this work suggest using the de Groot - van Leeuwen - van Weert (GLW) forms of the energy-momentum and spin tensors, together with their conservation laws, as the building blocks for construction of hydrodynamics with spin. The GLW framework follows naturally from the kinetic-theory considerations which directly show that the GLW hydrodynamic equations can be obtained as the zeroth and first moments of the Boltzmann equation --- both in the case of a semi-classical treatment of the Wigner function and in the case of classical description of spin. One has to admit, however, that at the moment only the boost-invariant solutions for such a scheme have been obtained, see Sec.~\ref{sec:appl}. A construction of new solutions needs much further research. 
	
	\item[{ 2.}] The GLW framework can be connected with the canonical one (obtained with the help of the Noether theorem from the underlying Lagrangian) through a pseudo-gauge transformation that has been explicitly constructed. Both, the GLW and canonical frameworks include spin degrees of freedom, hence can be used at the same footing to describe spin polarization phenomena. An advantage of the GLW formalism (i.e., the symmetric energy-momentum tensor and the spin tensor strictly conserved) is that it allows for a simple physics interpretation of the hydrodynamic variables and straightforward implementation of the initial conditions --- in the GLW case the latter are directly given by the values of the hydrodynamic variables, while for the canonical case they involve derivatives.  
	
	\item[{ 3.}] The pseudo-gauge transformation from the canonical form to the Belinfante one, performed at the level of hydrodynamic tensors, misses a dynamic equation for the spin degrees of freedom. This leads to a formalism that is not completely satisfactory for description of the polarization phenomena --- the total angular momentum in the Belinfante approach has the form of the orbital angular momentum which can be always locally set equal to zero by a Lorentz transformation. The information about spin polarization should be obtained by additional methods that go beyond the scope of hydrodynamics discussed herein.
	
	\item[{ 4.}] In the canonical, GLW, and Belinfante approaches, the semi-classical expansion of the Wigner function leads to the same, symmetric expression for the energy-momentum tensor in the leading order in $\hbar$. This expression, with the spin chemical potential consistently neglected, can be used in General Theory of Relativity which is a classical theory.  
	
	\item[{ 5.}] In the Belinfante case one can directly connect
	the spin polarization with thermal vorticity. This has now become a common procedure followed by different phenomenological analyzes of the $\Lambda$-hyperon polarization. Nevertheless, it is not obvious how to construct a hydrodynamic picture in this case, which would allow for space-time studies of polarization. The dependence of equilibrium distributions on the thermal vorticity is a reflection of non-locality, since thermal vorticity depends not only on the hydrodynamic variables but also on their gradients. 
	
	\item[{ 6.}] Different frameworks described in this work lead to a separate conservation of the spin tensor. This behavior may be traced back to the locality of the collision term, which implies the conservation of internal angular momentum in binary collisions. Departure from the locality assumption is most likely a necessary condition to include dissipative interactions which eventually could make the spin polarization tensor $\omega_{\mu\nu}$ equal to the thermal vorticity $\varpi_{\mu\nu}$ (as discussed in the point above).
	
	\item[{ 7.}] It was already pointed out by Hess and Waldmann in their seminal paper on the kinetic theory of particles with spin, Sec.~12 of \CIT{Hess:1966aa}, that locality of the collision operator makes their model unable to describe the Barnett effect, i.e., the orientation of spins by a local or uniform rotation of the system. Recent developments obtained within the Chiral Kinetic Theory \CITn{PhysRevLett.113.182302,Yang:2018lew} may give some insights how to improve the frameworks discussed in this work. 
	
	\item[{ 8.}] The magnitude of the polarization three-vectors obtained with the quantum distribution functions may be not bounded from above, which suggests that such distributions are appropriate only for small values of the spin chemical potential. In this case the GLW framework coincides with the approach where spin is treated classically. Within these two frameworks the normalization of the polarization and Pauli-Luba\'nski vectors is fixed, hence, any additional schemes to renormalize their magnitude seem to be not justified. 
	
	\item[{ 9.}] Using the classical concept of spin one can formulate a consistent framework of hydrodynamics with spin, which for small values of the polarization agrees with the approach using relativistic spin-density matrices. The classical-spin approach is free from the problems connected with normalization of the polarization three-vector and indicates that the hydrodynamic system becomes anisotropic if the spin densities are large. The classical approach allows also for the explicit definition of a conserved entropy current. The classical approach presented herein may be considered as complementary to more involved formalisms such as world-line approach \cite{Mueller:2019gjj}.  
	
	\item[{ 10.}] In the future practical applications, it would be useful to consider first the expressions linear in the polarization tensor. In this case one can solve first the standard hydrodynamic equations and the spin evolution can be later analyzed as a problem of the polarization dynamics in a given hydrodynamic background.

    \item[{11.}] The physics picture presented in this work is restricted to entropy-conserving, perfect fluid with spin. The inclusion of dissipation is an obvious demand for future studies. Some work in this direction has been already done \cite{Hattori:2019lfp}.	
	
\end{itemize}

{\bf Acknowledgements:} We thank Francesco Becattini, Krzysztof Golec-Biernat, Bengt Friman, Amaresh Jaiswal,  and Enrico Speranza for iluminating discussions and very fruitful collaboration. 




\bibliographystyle{elsarticle-num} 
\bibliography{spin-lit}





\end{document}